\documentclass[graybox]{svmult}

\usepackage{natbib}

\usepackage{type1cm}        
\usepackage{makeidx}         
\usepackage{graphicx}        
\usepackage{multicol}        
\usepackage[bottom]{footmisc}

\usepackage{newtxtext}       %
\usepackage{newtxmath}       
\usepackage{esvect}

\newcommand\fdg{\mbox{$.\!\!^\circ$}}%
\newcommand\fm{\mbox{$.\!\!^{\rm m}$}}%
\newcommand\farcs{\hbox{$.\!\!^{\prime\prime}$}}
\newcommand\arcsec{\hbox{$^{\prime\prime}$}}

\makeindex             

\begin{document}

\title*{Optical polarimetry: Methods, Instruments and Calibration Techniques}
\author{\bf Andrei Berdyugin, Vilppu Piirola,  \and Juri Poutanen}
\institute{Andrei Berdyugin, Vilppu Piirola  \at Department  of Physics and Astronomy, 20014 University of Turku, Finland\\ 
\email{andber@utu.fi, piirola@utu.fi}
\and Juri Poutanen \at Department  of Physics and Astronomy, 20014 University of Turku, Finland; 
Nordita, KTH Royal Institute of Technology and Stockholm University, Roslagstullsbacken 23, SE-10691 Stockholm, Sweden; 
Space Research Institute of the Russian Academy of Sciences, Profsoyuznaya str. 84/32, 117997 Moscow, Russia  \\
\email{juri.poutanen@utu.fi}
}
%
%
\maketitle

\abstract{In this chapter we present a brief summary of methods,
instruments and calibration techniques used in modern astronomical
polarimetry in the optical wavelengths. We describe the properties
of various polarization devices and detectors used for optical
broadband, imaging and spectropolarimetry, and discuss
their advantages and disadvantages. The necessity of a proper
calibration of the raw polarization data is emphasized and methods
of the determination and subtraction of instrumental polarization
are considered.
We also present a few examples of high-precision measurements of optical polarization of black hole X-ray binaries and massive binary stars made with our DiPol-2 polarimeter, which allowed us to constrain the sources of optical emission in black hole X-ray binaries and measure orbital parameters of massive stellar binaries. 
}

\section{Introduction}
\label{sec:1}

The first measurements of polarization
(in the optical wavelengths)\footnote{The terms \emph{``optical light''} or \emph{``optical wavelengths''} are usually understood as the wavelength range from 300 to 1000\,nm. 
This is the range covered with the widely used UBVRI Johnson/
Cousins photometric systems. } were performed by Dominique Francois
Jean Arago and they date back to the first half of the 19th century. 
However, optical polarimetry became a mainstream method of
astrophysical observations only in the end of 1960s. This happened
due to the grown use of sensitive photoelectric detectors, such as
photomultiplier tubes (PMTs) and charge coupling devices (CCDs) in
optical photometry and spectroscopy.

\begin{figure}[b]
\sidecaption
\includegraphics[scale=.65]{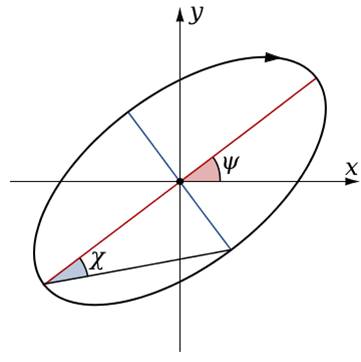}
\caption{Polarization ellipse of fully elliptically polarized
monochromatic electromagnetic wave. This ellipse is swept by the tip
of the electric vector $\vec{\it E}$ on the plane which is orthogonal to
the direction of propagation. Note: in radio astronomy the magnetic
vector $\vec{\it B}$ is chosen for characterization of polarization.}
\label{fig:1}   
\end{figure}

PMTs and CCDs are able to provide a sufficient S/N for measuring
optical polarization even for low fluxes received from astronomical objects. 
CCD cameras, which are \emph{panoramic} or
\emph{multi-cell} detectors, are capable of measuring polarization
in many elements of image or spectrum simultaneously. With the
introduction of the new detectors, powerful polarization devices and
new techniques of polarization measurements have been developed.
Currently, optical polarimetry is a well-established observational
tool, widely used on small and large optical telescopes for studying
a large variety of astrophysical objects, from planets to active galaxies. 
As also emphasized at the EWASS 2018 meeting, 
about 4000 refereed publications on optical polarimetry have been published since year 2000.

\subsection{Description of Polarization: Stokes Parameters, Linear and Circular Polarization}
\label{subsec:1.1} 

Polarization can be described by the \emph{normalized Stokes
parameters} $q$, $u$ and $v$ defined as:
\begin{equation}
q = Q / I\;, \;\; u  = U / I\;, \;\;v = V / I. 
\end{equation}
Here \emph{Q, U} and \emph{V} are the \emph{absolute Stokes parameters}:
\begin{equation}
 Q = I_{P} \cos2\psi \cos2\chi \; , \;\;
 U = I_{P} \sin2\psi \cos2\chi \;,  \;\;
 V = I_{P} \sin2\chi \;, 
\end{equation}
${I}$ is the total intensity of the partially polarized light and
$I_{P}$ is the intensity of its \emph{elliptically} polarized
fraction. The angle $\psi$, which is also often denoted in the literature as
$\theta$ (Theta) or PA (Polarization Angle), is the \emph{azimuthal}
angle of the \emph{polarization ellipse} (see Fig.~\ref{fig:1}). In
astronomical polarimetry it is measured counter-clockwise from the
direction to the celestial North in the equatorial coordinate
system. The angle $\chi$ is a measure of the \emph{eccentricity} of
the polarization ellipse, so that $\tan\chi$ is the ratio of its
minor and major axes.  From the normalized Stokes parameters the
\emph{degree} $P_{\rm L}$ (or ${\rm PD}_{\rm L}$) and \emph{direction} $\theta$ (PA) of linear polarization, and the degree of circular polarization, $P_{\rm C}$ (or ${\rm PD}_{\rm C}$), can be
found:
\begin{equation}
  P_{\rm L}= \sqrt{q^2 + u^2} \;  , \; \;  \theta = \frac{1}{2} \arctan{\frac{u}{q}} \; , \;\; P_{\rm C} = v.
\end{equation}

\subsection{Mechanisms Producing  Polarization of Astrophysical Objects}
\label{subsec:1.2}

There are several physical processes which may be responsible for
producing substantial amount of linear  and/or circular
polarization in the optical part of the spectrum. Those processes are:
\begin{enumerate}
\item{Reflection of light from a solid surface (e.g. the surface of a planet or asteroid).}
\item{Scattering of light on free electrons, atoms, molecules or dust particles.}
\item{Propagation of light in the presence of the magnetic field (Zeeman and Paschen-Back effects).}
\item{Cyclotron and synchrotron radiation of accelerated free electrons moving in the magnetic field.}
\item{Interstellar polarization of distant stars due to optical dichroism of non-spherical interstellar dust grains aligned by interstellar Galactic magnetic field.}
\end{enumerate}
Through the processes of light emission, reflection, scattering and
absorption, optical polarimetry can provide us with important
information on the properties of matter, such as spatial
distribution, density, particle composition and size. It also helps us 
to derive the properties of stellar and interstellar magnetic
fields, e.g. the field strength and geometry. 

\section{Polarization Modulators, Analyzers and Detectors for Optical Polarimetry}
\label{sec:2}

Most of optical polarimeters employ a polarization \emph{modulator}
and a polarization \emph{analyzer}. Polarization modulator is the
optical device which \emph{modulates} the state of polarization of
the incoming light beam with a given frequency. Polarization
analyzer is the unit which separates the two orthogonally polarized
components of light for measuring their intensities with the
\emph{detector}, synchronized with the modulation frequency.

\subsection{Modulators: Retarders with Constant Phase Shift (Wave Plates)}
\label{subsec:2.1}

A \emph{wave plate  } is a plane parallel plate made of a uniaxial
crystal cut in the way that its optical axis (the direction in which 
refraction index $n_{e}$ is minimum) is parallel to the plate surface. 
The light polarized \emph{along} this axis moves \emph{slower} than the 
light polarized in the orthogonal direction. Such plate introduces a phase shift, i.e., \emph{retardation} $\tau$ between the orthogonally polarized
components of the light beam entering the plate \citep{Serkowski74}: 
\begin{equation} 
\tau = 2\pi \Delta / \lambda, \quad \Delta= (n_{e} - n_{o}) s, 
\end{equation}

where $s$ is the thickness of the retarder, $\lambda$ is the
wavelength, $n_{e}$ and $n_{o}$ are the refractive indices for
the light components polarized parallel and perpendicular to the
optical axis of the crystal, respectively, and  $\Delta$ is the \emph{path difference}.

There are two types of retarders widely used in optical polarimetry:
a retarder with $\tau = \pi/2$ (or $\Delta= \lambda/4$) is called
\emph{quarter wave plate} (QWP), and with the $\tau = \pi$ (or
$\Delta= \lambda/2$), called \emph{half wave plate} (HWP). The QWP changes
the circularly polarized light with $I = V$ into linearly polarized
light with $I = (Q^{2} + U^{2})^{1/2}$. The HWP \emph{rotates} the
plane of polarization of the linearly polarized light beam. Thus, if
$\phi$ and $\theta$ are the directions of HWP axis and polarization
of the incoming light beam measured with respect to the northern
celestial pole, the direction of polarization plane of the beam
emerging from the HWP becomes $2\phi - \theta$.

For a good optical retarder, the retardance should change as little
as possible over the optical wavelengths. Currently available
\emph{superachromatic} optical wave plates are close to this
requirement. For example, the multi-component superachromatic half and
quarter wave plates offered by Bernhard Halle have the maximum
deviation of retardation $\pm 1.3\%$ (HWP) -- $\pm 4\%$ (QWP) and
variation of the axis direction $\pm 1.5^\circ$ (QWP) -- $\pm
2.0^\circ$ (HWP) over the spectral range 310--1100\,nm. They are
made of quartz and magnesium fluoride and available in various
apertures, up to 50x50 mm. Polymer-made superachromatic wave plates
offered by Astropribor closely match the optical properties of the
best crystal-made retarders, although over the slightly narrower
wavelength range. They are more durable  and available in a larger
set of apertures, up to 60 mm.

A QWP rotated in steps of $90^{\circ}$ and  placed in front of a
polarization analyzer is frequently used to measure circular
polarization $v$. 
A HWP rotated by steps of 22\fdg5  is used for
linear polarization (Stokes $q$ and $u$) measurements (see Section \ref{subsubsec:3.1.1}). 
Moreover, a QWP rotated with the steps of 22\fdg5 can
be used to measure all Stokes parameters $q$, $u$ and $v$. However,
in the latter case the intensity modulation of the light beam due to
linear polarization has twice smaller amplitude, which reduces the
efficiency of linear polarimetry by 50\% \citep[see e.g.][]{Serkowski74}.

\subsection{Modulators: Retarders with Variable Phase Shift (PEMs and FLCs)}
\label{subsec:2.2}

\emph{Photoelastic or piezoelectric modulator} (PEM) is made of
non-birefringent slab of material which is stressed using
piezoelectric transducer at the natural slab frequency $f_{0}$. This
allows to greatly reduce the power needed to sustain the standing
wave in the slab. PEM operated at the base frequency $f_{0}$ acts as
a $variable$ QWP introducing the periodic phase shift $\lambda/4$
between orthogonally polarized components of the incoming light
beam. Placed in front of the analyzer, PEM can be used for direct
measurement of circular polarization. If the PEM is driven at the
frequency $2f_{0}$, the phase shift $\lambda/2$ is introduced and
linear polarization can be directly measured. However, in the latter
case, either the PEM itself or the lower section of polarimeter
(below the PEM) must be rotated to two orientations separated by
$45^{\circ}$ to allow measurement of the complete set of Stokes $q$
and $u$ \citep[see][]{Kemp81,Hough}.

Nominal modulation frequencies of PEM are tens of kHz, well above
fluctuations produced in the Earth's atmosphere which makes it well
suitable for high-precision polarimetry. Among other advantages are
the large useful aperture and possibility to use it in the fast
optical beams (up to cone angles of $\leq 50^{\circ}$). The PEMs for
optical polarimetry are usually made of fused silica with the anti-reflecting 
coating optimized for the desired photometric passband.

\emph{Ferro-electric liquid crystal} (FLC) modulators are made of a
thin layer of liquid crystal material, sandwiched between two glass
plates. FLCs used in optical polarimetry have a fixed $\lambda/2$
retardation with switchable orientation of the optical axis
controlled by applied drive voltage. Thus, they can be used for
measuring linear polarization in a way similar to that with the
PEMs. The main advantage of the FLC modulator over the PEM is the
nearly square-wave modulation, which makes it more efficient.

FLCs are operated at modulation frequencies in the range from
hundreds of Hz to a few kHz which is fast enough to get rid of
effects due to atmospheric intensity fluctuations. The FLCs are
temperature sensitive devices and must be used in a temperature
controlled environment to provide internal stability of the
switching rate and the switching angle. The FLCs are also known for
producing a large instrumental polarization from multiple internal
reflections in the birefringent material between the plates. The
magnitude of this polarization can be in the range of 0.1--0.3\%
\citep{Bailey15} and is highly variable with the wavelength. Special measures
must be taken to properly calibrate and remove it.

Similarly to PEMs, the FLCs currently available on the market are
optimised for different spectral regions. However, unlike the
constant phase-shift wave plate which can be made superachromatic,
the change of passband from blue to red in the polarimeter equipped
with a variable phase-shift retarder requires changing the modulator
unit.

\subsection{Analyzers for Optical Polarimetry: Single and Double-beam Units}

\label{subsec:2.3} A single \emph{polaroid} is probably the very
first analyzer ever used in optical polarimetry. A simple polaroid
is made of hematite crystals, embedded in a polymeric transparent
thin film, or layered on a glass surface. Crystals are aligned in
one direction. Polaroid possesses a property which is called
\emph{optical dichroism}: a strong absorption of light linearly
polarized in the direction parallel to the orientation of crystal
axis. Thus a rotated polaroid can be used as the simplest
\emph{single-beam} analyzer for linearly polarized light.

Obvious disadvantage in using polaroid is that half of the intensity
of the incoming light beam is lost. Because in the most astronomical
polarimetry applications the accuracy is determined by the number of
available photons, it is important to retain as much light as possible. 
For this reason, polaroids and other single-beam analyzers (i.e.
Nicol and Glan-Thompson prisms) are rarely employed for optical
polarimetry presently.  However, in certain cases, the usage of
polaroids can be justified, e.g. for the imaging polarimetry of
extended objects (reflecting nebulae) as, for example, is done in the
case of HST polarimeter.

A \emph{double-beam} analyzer, or \emph{polarization beam-splitter},
has the advantage of splitting the incoming light beam into two
orthogonally polarized beams in such a way that both of them can be
measured simultaneously. Such analyzer utilizes the property found
in some uniaxial crystals, which is called a \emph{birefringence}. In
these crystals the orthogonally polarized components of the
electromagnetic wave travel with different speed, i.e. have
different refractive indices. The birefringence degree $\Delta n$ is
defined as a difference between the refraction indices of the
$extraordinary$ and the $ordinary$ rays $n_{e} - n_{o}$. The
extraordinary (e-ray) is polarized in the direction parallel to the
crystal \emph{principal axis}, and the ordinary (o-ray) is polarized
in the perpendicular direction. There are crystals with negative
$\Delta n$ (beryl, calcite) and positive $\Delta n$ (magnesium fluoride, quartz).

\subsubsection{Double-beam Polarization Analyzers for Optical Polarimetry: Plane-parallel Calcite Plate, Savart Plate and Wollaston Prism}
\label{susubsec:2.3.1} 

Because a double-beam polarization analyzer is the key element in
many existing optical polarimeters, is worth to mention the most
important requirements which a good optical polarization
beam-splitter should meet:
\begin{enumerate}
\item{High transparency in the optical spectral range.}
\item{Constant birefringence $\Delta n$ over the optical wavelengths.}
\item{No aberrations are introduced to the telescope exit pupil.}
\item{The same optical path length for the e and o-rays.}
\item{High polarization efficiency, i.e. the completely non-polarized light beam is split in two
100\% orthogonally polarized beams with intensities $I_{e}$ and $I_{o}$ so that $I_{e} = I_{o}$.}
\end{enumerate}

\begin{figure}[t]
\sidecaption
\includegraphics[scale=0.1]{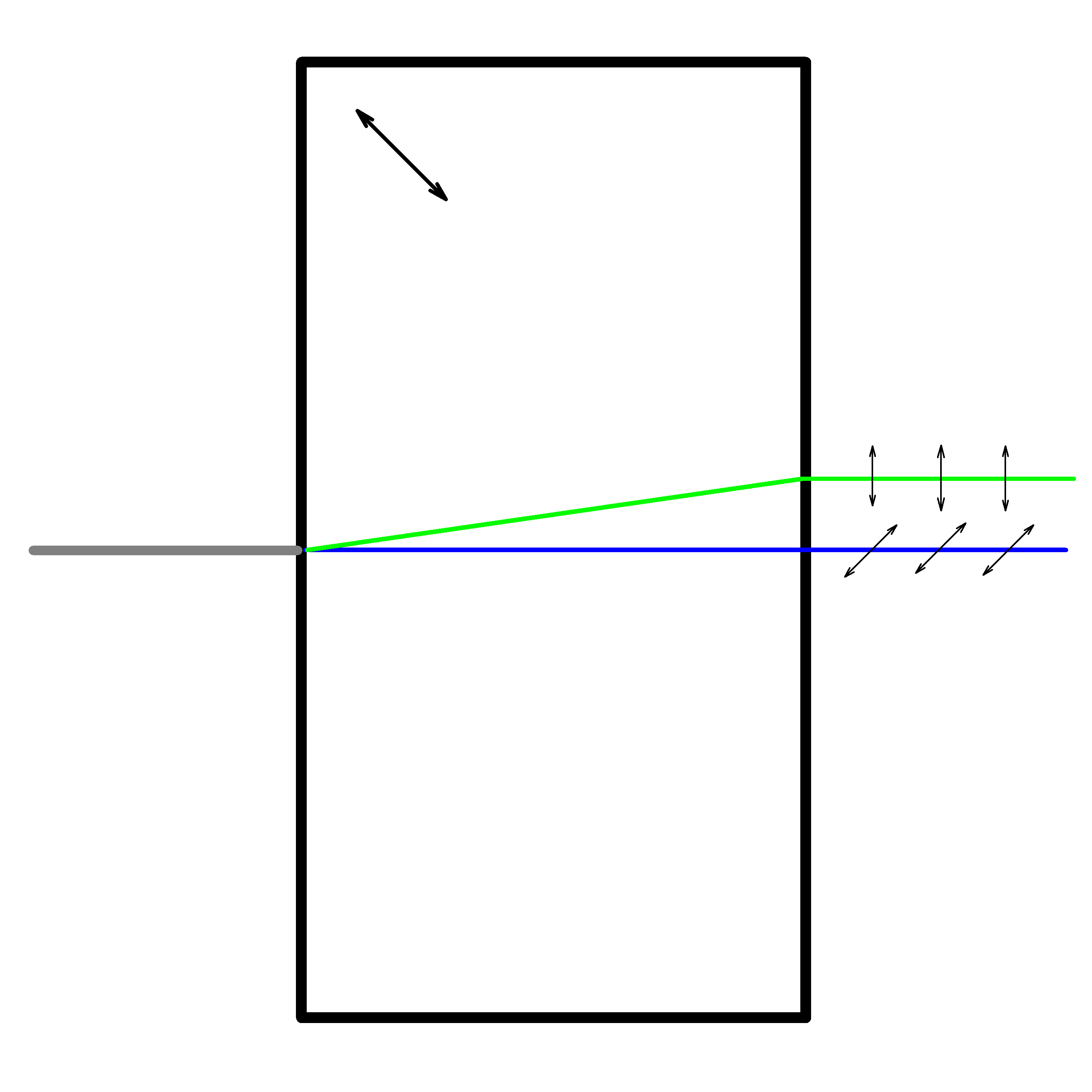}
\caption{Plane-parallel calcite plate. The principal axis has the angle of $45^{\circ}$ to the plate normal. The direction of travel of the o-ray (blue) is not changed, but the e-ray (green) is shifted in the parallel direction.}
\label{fig:2}       
\end{figure}

There are three types of double-beam polarization analyzers most frequently used in optical broadband and spectropolarimetry.
The most simple double-beam polarization analyzer is a
\emph{plane-parallel calcite plate}. It is made of calcite crystal
cut parallel to the plate of cleavage (Fig.~\ref{fig:2}). Such plate
splits stellar image in two by introducing a shift of the e-ray. The
image separation (for $\lambda = 550$ nm) is $d = 0.109 \times h$,
where $h$ is the calcite thickness. Being simple, this polarization
device can be manufactured up to the required thickness and aperture
at a reasonable cost. Because it is a single-piece unit, it is not
so fragile in comparison with other devices and can be handled with
less caution. However, the optical path for the e and o-rays is not
the same, and both images cannot be focused precisely on the same
plane. Moreover, the light losses by reflection on calcite surface
for the e and o-rays are different. Single calcite plate introduces
a noticeable \emph{dispersion} into the e - stellar image in the
blue -- near UV wavelengths. Despite of these shortcomings, the
plane-parallel calcite plate is often used in applications where the
image quality is not critical.

\begin{figure}[b]
\sidecaption
\includegraphics[scale=0.7]{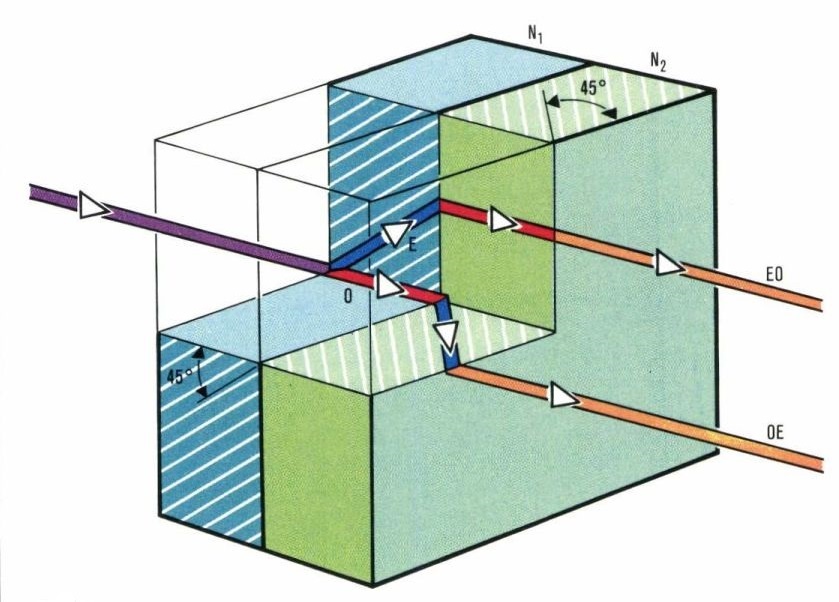}
\caption{Savart plate. When entering into the second crystal, the
o-ray produced by first one becomes the e-ray and is displaced in
the direction perpendicular to the first beam displacement. The
result is two rays displaced along a diagonal. The optical path
difference for the o- and e-rays is zero for normal incidence.}
\label{fig:3}       
\end{figure}

Most of the inconvenient features of a single calcite plate are
avoided in \emph{Savart plate} which is made of two plane-parallel
calcite plates cemented with their principal axes at $45^{\circ}$ to
the surface normal and rotated through $90^{\circ}$ with respect to
each other (Fig.~\ref{fig:3}). Savart plate seems to be the best
analyzer for the stellar polarimetry with panoramic (i.e. CCD
camera) detectors. Nevertheless, accurate measurements of
polarization might not be possible in crowded regions of the sky
due to overlapping of the e- and o-images of the target and the
nearby field stars. This problem can be solved by mounting the
polarimeter on a rotator which can be turned at certain angle to
change the direction of the stellar images split on the sky in 
a way to avoid the overlapping. Another solution is to put a stripe mask
in the focal plane of the telescope, but this requires using
re-imaging optics (collimator and camera lenses).

\begin{figure}[t]
\sidecaption
\includegraphics[scale=.70]{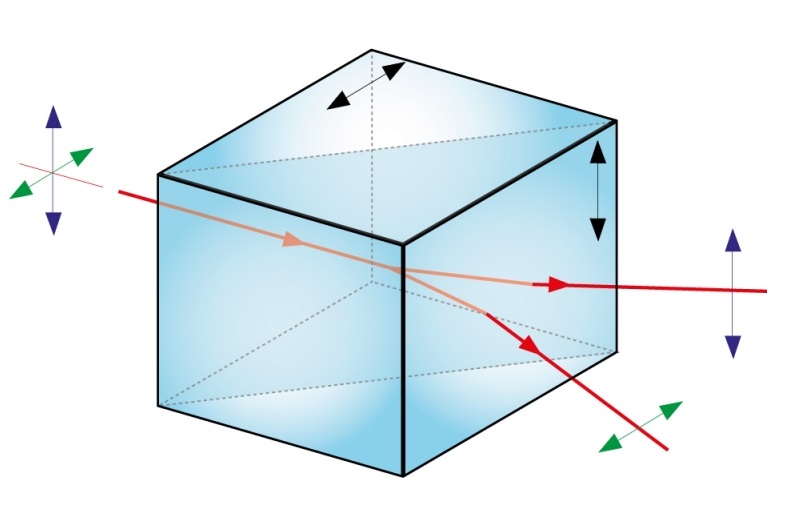}
\caption{Wollaston prism made of two cemented calcite wedges with
orthogonally directed optical axes. The angle of separation of the
o- and e-rays can be within $5^{\circ}  - 20^{\circ}$.}
\label{fig:4}       
\end{figure}

If polarimetry is done with photomultipliers, a large separation
between the beams is convenient. The best double-beam analyzer which
provides a large image separation and high efficiency is the
\emph{Wollaston prism}. The \emph{two-wedged} Wollaston prism
(Fig.~\ref{fig:4}) is made of two cemented (calcite) or optically
contacted (quartz, magnesium fluoride) wedges with optical axes
orthogonal to each other and perpendicular to the direction of
propagation of the incoming light beam. The angular separation
between the outgoing e- and o-rays in the Wollaston prism depends on
the number of wedges $N$, construction angle of the outer wedges $A$
and birefringence of material as:
\begin{equation}
 \alpha \simeq 2 \arctan[(N - 1)\ |n_{o} - n_{e}|\ \tan A] \;.
\end{equation}

In the optical wavelengths, calcite is the preferred material,
because it is transparent from 350 to 2200 nm and has a high
birefringence. Thus, two-wedged calcite Wollaston prisms offered by
Bernhard Halle, can provide the beam separation up to $20^{\circ}$.
If a higher beam separation is required, the \emph{three-wedged}
calcite Wollaston prism can be used. It provides image separation up
to $28^{\circ}$ and gives the best image quality, approaching
closely the ideal analyzer for the optical polarimetry \citep{Serkowski74}. 
If a wider spectral range is needed, i.e. with the near UV and near-IR
coverage, quartz or magnesium fluoride must be used. However, due to
the much lower birefringence of these crystals, the beam separation
is by an order of magnitude smaller. In comparison with the single
calcite plate, the Wollaston prism is a more complex and expensive
optical device and must be handled with care.

\subsection{Detectors for Optical Polarimetry: CCDs, PMTs and APDs}
\label{subsec:2.4}

There are three types of detectors used in optical polarimetry: CCD
cameras, photomultiplier tubes (PMTs) and avalanche photodiodes
(APDs). Each type of detector has its own advantages, disadvantages
and areas of applications. There are also emerging
detectors for the optical and NIR spectral ranges, namely
MKIDs and Saphira APD arrays. These are likely to play an
important role in future polarimeters, particularly Saphira 
based systems for NIR \citep{Finger14}.

\subsubsection{Detectors for Optical Polarimetry: CCD Cameras}
\label{subsubsec:2.4.1}

CCD cameras are currently the most widely used detectors for
photometry and spectroscopy at the optical wavelengths. Among their
main advantages are the high quantum efficiency (QE), great versatility
and convenience of data acquisition. Moreover, being panoramic
(multi-cell) registration devices, they are very well suited for
spectroscopic and imaging applications. This feature is also very
useful for \emph{double-beam polarimetry}, because it allows us to
record two orthogonally polarized stellar images or spectra
simultaneously on the same detector, so that variations in
atmospheric seeing and transparency will have exactly the same
effect on both. Moreover, the intensity of background sky is
registered automatically on the pixels surrounding stellar images
and this eliminates a necessity to measure intensity and
polarization of the sky separately.

\begin{figure} 
\includegraphics[width=0.9\columnwidth]{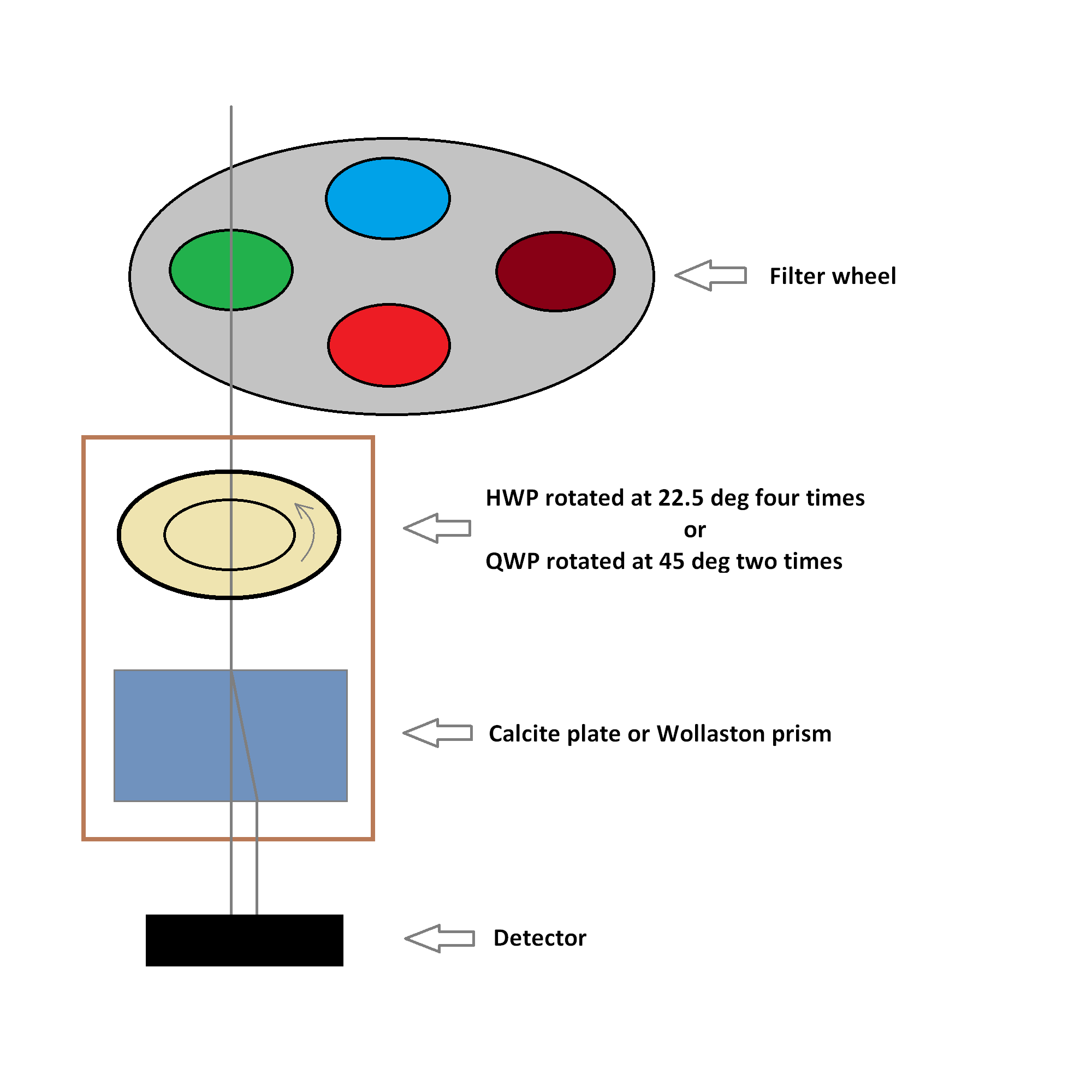}
\caption{The layout of the CCD optical polarimeter. Polarization
components can be installed between the CCD detector and filter
wheel. After passing the beam-splitting analyzer, the incoming light
ray is split parallel in two orthogonally polarized rays. Instead of
plane-parallel calcite shown on the figure, the other analyzer, e.g., Wollaston prism can be used.}
\label{fig:5}       
\end{figure}

The CCD cameras employed in optical polarimetry can be roughly
divided into two classes: smaller units with the thermoelectric
cooling and pixel area up to about 2048$\times$2048 (used mostly for
broadband polarimetry) and large pixel area (or mosaic) units with
cryogenic cooling (used for spectropolarimetry, imaging polarimetry
and in multi-functional instruments with optional \emph{``polarimetry
mode''}). The cameras of the first class can be connected to the
computer via standard USB port used for the camera control and data
acquisition. They are often equipped with the high-grade, low noise
CCD detector with sensitivity optimized either for the blue-visual
or visual-red spectral ranges with the maximum QE $\geq 90\%$. Many
such cameras are available currently on the market at a reasonably
low cost. Thus, they are well suitable for building a low-budget
instrument for stellar polarimetry at the optical wavelengths. The
principal design of such CCD polarimeter is shown in
Fig.~\ref{fig:5}.

In broadband CCD polarimetry the data reductions can be done with
\emph{aperture photometry} method which normally includes
calibration (subtraction of dark, bias and flat-fielding), sky
subtraction, determination of intensities of e- and o-stellar images
and converting them to the \emph{intensity ratios} $Q_{i} = I_{e}/I_{o}$. 
The $Q_{i}$ must be determined for the each orientation of the wave plate $i = 0^{\circ}, 22\fdg5, 45^{\circ}, ...$ 
The data reduction in CCD spectropolarimetry is done
in the similar way: the e- and o-spectra must be extracted and, after wavelength calibration, the intensity ratios $Q_{i}(\lambda)$ are computed. 
The Stokes parameters of linear or circular polarization are obtained from the intensity ratios $Q_{i}$ with the formulae given in Section \ref{subsubsec:3.1.1}.

The CCD detectors, while being convenient for optical polarimetry, have also some shortcomings, e.g. the relatively small full-well capacity ($ \leq 200000 e^{-}/$pixel) which results in fast pixel saturation (overexposure) of the images of the bright stars. 
The image download time for the small-sized CCD camera is typically around of $\simeq$1--2\,s, while for the large, mosaic-type CCD detector it can be in range from few tens of seconds up to $\simeq$1\,min. 
Thus, CCD polarimetry of bright targets, necessarily
done with short exposures, may become very inefficient, because most of the time is spent on image downloads.

The download time problem is removed in CCD cameras with \emph{fast frame transfer}. The CCD chip in such a camera has two areas, one is for the image exposure and another for the temporal image storage. As soon as the exposure is
taken, the image is swiftly shifted vertically from the exposure area into the masked storage area from where it is transferred to the camera readout register. Thus, for example, ANDOR iXon Ultra 897 EMCCD camera can provide an impressive frame rate up to 56 full-frame images/sec. Even higher frame rate is achieved in the sub-frame readout mode. The frame-transfer EMCCDs are
rather expensive and available only with a relatively small pixel
area, up to 1024$\times$1024. Nevertheless, such camera can be considered
as a very good option for building an efficient and versatile CCD
polarimeter operating in the optical wavelengths.

\subsubsection{Detectors for Optical Polarimetry: Photomultiplier Tubes}
\label{susubsec:2.4.2}

Although most of the existing astronomical polarimeters employ CCD
cameras, \emph{photomultiplier tubes (PMTs)} are still used in
optical polarimetry. The main advantages of PMTs, making them a
good choice as detectors for certain applications, are the
instant readout, wide dynamical range and ability to faithfully
register high fluxes. Thus, a typical PMT retains linear response up
to the level of $10^{6} e^{-}$/s in the pulse counting mode and up
to $\leq 10^{8} e^{-}$/s in a direct photo-current amplification
mode.

Polarimetry is a ``photon hungry'' observational technique. The
accuracy is critically dependent on the amount of registered photons
or ADUs as:
\begin{equation}
 \sigma_{q, u, v} = k N^{-1/2} ,
\end{equation}
where $k$ is the analyzer efficiency and $N = N_{e} + N_{o}$ is the
total number of registered ADUs or counts. Thus, in order to achieve
the precision at the level of $10^{-6}$ (part per million or ppm),
$10^{12}$ ADUs must be collected. The early pixel saturation, a
bottleneck for the CCD polarimeter when it is used for observation
of the bright targets, normally is not an issue for the polarimeters equipped
with a PMT detector. Moreover, the ``instant'' readout allows one to use
PMT with the fast polarization modulator such as PEM or FLC. Not
surprisingly, the PMT detectors have been used in the most precise
polarimeters such as HIPPI \citep{Bailey15} and POLISH2 \citep{Wiktorowicz15}.

The major disadvantage of the PMT detectors is their relatively low
quantum efficiency which is typically in the  range 10--30\%. 
The best in their class, super bialkali (SBA) and ultra bialkali
(UBA) PMTs have QE $\simeq 40\%$ with the peak near 400 nm. Thus, the polarimeters with PMT detectors, while being able to achieve an accuracy level of few ppm for bright stars, still cannot compete with the CCD polarimeters in the low flux domain. Because the PMT is a single-cell detector, it is
suitable only for single target polarimetry of point-like objects.
All existing PMT polarimeters utilize a diaphragm in the focal plane
of the telescope, where the measured star is placed. The intensity
and polarization of sky background are measured separately by either
pointing the telescope away from the star to the clear area of the sky
or with a separate sky registration channel.

\subsubsection{Detectors for Optical Polarimetry: Avalanche Photo-Diodes}
\label{subsubsec:2.4.3}

The \emph{avalanche photo-diode} or APD is the semiconductor analog
to the PMT. In an APD, as with any other photodiode, incoming
photons produce electron-hole pairs; however, the APD is operated
with a large reverse \emph{bias voltage} (up to 2 kV), which
accelerates photoelectrons. Those electrons collide with atomic
lattice, releasing additional electrons via secondary ionisation.
The secondary electrons are also accelerated, which results in an
\emph{avalanche} of charge carriers, hence the name.

Like the PMTs, APDs can work with high modulation frequencies (up to 
a few MHz) and can tolerate even higher fluxes, up to $\leq 10^{9}
e^{-}/$s. They have a higher QE (up to 80\% in the near IR) in
comparison with the PMTs, but smaller or comparable QE in the blue
wavelengths. The APDs have been used as detectors in PlanetPol \citep{Hough}, POLISH  \citep{Wiktorowicz08} and OPTIMA \citep{Kanbach08} polarimeters. Although the APD detectors may be better suitable for high-precision polarimetry of the bright targets in the near IR wavelengths, they do not offer significant advantage over the PMTs in the blue and visual spectral ranges. In comparison with the PMT (and CCD) detectors, the APDs have a higher dark current and are more noisy. Another disadvantage is the small effective
photosensitive area, typically a few mm$^{2}$, which requires a very
precise positioning of the telescope exit pupil on the detector
surface.

It seems that APDs are becoming less popular detectors for high-precision optical polarimeters. The HIPPI (a successor to PlanePol) and POLISH2 (a successor to POLISH) are both employing PMTs.

\section{Broadband, Imaging and Spectropolarimetry in the Optical Wavelengths}

The various methods of measuring polarization in optical wavelengths
can be divided into three distinct groups: \emph{broadband
polarimetry}, i.e., polarimetry of stellar-like objects with the use
of broadband optical filters; \emph{imaging polarimetry} of extended
objects, like nebulae, usually made also with filters; and
\emph{spectropolarimetry}, i.e., measuring polarization over the
regions of the optical spectrum with certain \emph{spectral
resolution}. Imaging polarimetry can be also done for the stellar-like
targets. This method of observations has the advantage of simultaneous 
registration of sky background and comparison object which effectively 
shortens any exposure time. 

\subsection{Instruments for Broadband Optical Polarimetry}
\label{subsec:3.1}

There are many instruments in this group utilizing different
approaches and focused on achieving different goals. We give here a
brief description of the most commonly used techniques and designs
currently employed in optical broadband polarimetry.

\subsubsection{Double-image CCD polarimeters}
\label{subsubsec:3.1.1}

The layout of the most commonly used broadband optical CCD
polarimeter is shown in Fig.~\ref{fig:5}. It consists of four parts:
1) detector (CCD camera); 2) filter wheel; 3) rotatable half wave
(HWP) or quarter wave plate (QWP); and 4) double-beam polarization
analyzer such as plane-parallel calcite (or Savart) plate or
Wollaston prism. The rotation of the HWP (QWP) is normally done with
the use of a high-precision \emph{stepper motor}.

In order to measure \emph{linear polarization}, a series of CCD
images must be taken at the four positions of the HWP: $0^\circ$,
$22\fdg5$, 45$^\circ$ and $67\fdg5$. Next step is to measure the
intensities of the two orthogonally polarized stellar images $I_{e}$
and $I_{o}$ and compute their ratio $Q = I_{e}/I_{o}$. Then the
normalized Stokes parameters $q$ and $u$ can be computed as:
\begin{eqnarray}
Q_{m} &=& Q_{0} + Q_{22.5} + Q_{45} + Q_{67.5}, \\
q &=& (Q_{0} - Q_{45})/Q_{m}, \\
u &=& (Q_{22.5} - Q_{67.5})/Q_{m} .
\end{eqnarray}

\emph{Circular} polarization can be measured with the QWP in the
following way: two CCD images are taken at the positions of
$45^\circ$ and $135^\circ$ (or $-45^\circ$ and $45^\circ$) and the
degree of circular polarization $P_{\rm C} = v$ is computed as:
\begin{eqnarray}
Q_{m} &=& Q_{45} + Q_{135},  \\
v &=& 0.5 \  (Q_{45} - Q_{135})/Q_{m} . 
\end{eqnarray}

\begin{figure}[b]
\includegraphics[scale=1.0]{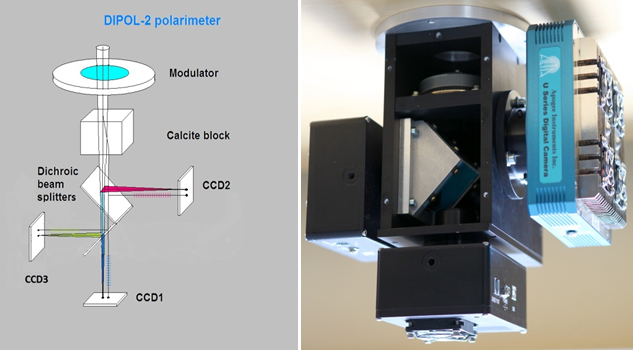}
\caption{Left: Schematic layout of the main components of the
DiPol-2 polarimeter. Right: photo of DiPol-2 taken with the front
cover removed. Polarization unit consisting of calcite, exchangeable
retarder and stepper motor is installed in the upper section.
Dichroic mirrors are installed in the lower section. Apogee Alta U47
CCD (B-band) is on the right side, two ST-402ME CCDs are on the left
(R-band) and on the bottom (V-band).}
\label{fig:6}       
\end{figure}

The setup shown in Fig.~\ref{fig:5} can be used as a polarimetric
``add-on'' to a CCD photometer, a low resolution spectrograph or a
multi-functional instrument. The rotated wave plate and two-beam
polarization analyzer can be moved into the optical path and removed
when they are not necessary. This method of measurements, being
simple, is probably the most efficient for the broadband and
low-resolution spectropolarimetry of faint point-like sources. 
Simultaneous registration of two polarized images (or spectra) helps to maximize efficiency and to avoid most of systematic errors.
Currently it is used in several existing instruments, e. g., ALFOSC at the NOT, ORM (La Palma) and EFOSC and FORS2 at the ESO (Chile).

The layout shown above can be easily expanded and enhanced. 
For example, the DiPol-2 polarimeter shown in Fig.~\ref{fig:6} does not have a filter wheel, but employs two dichroic beam-splitters installed after the
polarization unit. The light beam is split into three passbands (B,
V and R) registered with three CCD cameras simultaneously. Because
DiPol-2 uses a plane-parallel calcite plate as a polarization
analyzer, without having a focal plane mask, the o- and e-components
of the sky overlap everywhere on the CCD image. Thus, the sky is not
split into polarized components and sky polarization automatically
cancels out when the total sky intensity $I_{e} + I_{o}$ is
subtracted from the stellar images in data reductions \citep{Piirola14}. DiPol-2 has been built in three copies and used for high-precision polarimetry 
of the low-mass X-rays binaries \citep{Kosenkov,Veledina} (see Sect.~\ref{subsec:blackholes}), early-type binaries \citep{Berdyugin16,Berdyugin18} (see Sect.~\ref{subsec:earlytype}),  
novae \citep{Harvey} and study of the interstellar polarization in the vicinity of the Sun \citep{Frisch}. DiPol-UF, a successor to DiPol-2, currently being built at the University of Turku, employs three ANDOR iXon Ultra 897 EMCCD 
cameras as detectors and has a retractable polarization unit which 
allows operation of the instrument also as a fast three-band photometer 
(simultaneous in BVR).

\subsubsection{High-Precision Double-Image Polarimeters with High-Frequency Modulators}
\label{subsubsec:3.1.2}

\begin{figure}[b]
\sidecaption
\includegraphics[scale=.60]{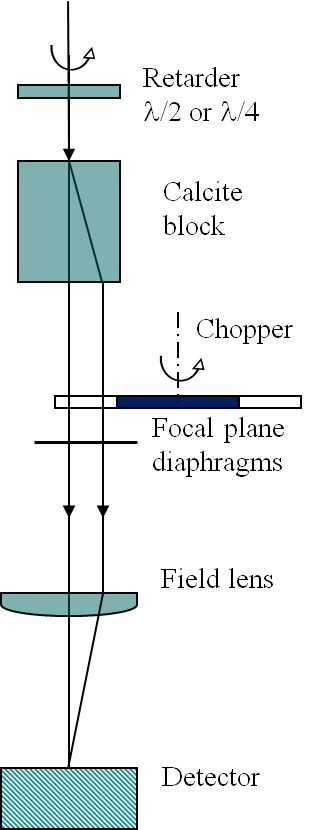}
\caption{The schematic layout of the TurPol polarimeter. The rotating
chopper is installed after calcite block and in front of one of
the selectable double diaphragms placed in the diaphragm wheel.
After the double diaphragm, the field lens forms the exit pupil on
the PMT detector. The registration of the intensities of e- and o-images of the star by the PMT is synchronized with the chopper
rotation. In the real instrument, the light is split in five (U, B,
V, R and I) beams by the four dichroic mirrors installed after the
diaphragm section and registered simultaneously with the five PMTs.}
\label{fig:7}       
\end{figure}

There are several examples of currently existing instruments
employing the techniques of fast modulation. One example is the
UBVRI polarimeter \emph{TurPol}  \citep{Piirola73,Piirola88},
the principal layout is shown in Fig.~\ref{fig:7}. In
TurPol, the modulation of the incoming light, split by the
calcite plate into two parallel beams, is done with a rotating
mechanical chopper. The frequency of modulation is 25 Hz, which is
enough to eliminate most of atmospheric effects. TurPol is equipped
with a discretely rotated interchangeable $\lambda/2$ and
$\lambda/4$ wave plates and is able to measure linear or circular
polarization of the stellar-like objects with an accuracy up to a
few times per $10^{-5}$ in the five UBVRI passbands
simultaneously. The effect of sky polarization is cancelled out due
to the overlapping of the orthogonally polarized sky images in each
of the two diaphragms. The total sky intensity $I_{e} + I_{o}$ is
measured separately by pointing the telescope at the nearby clear
area of the sky. TurPol has been made in several
copies and still used for observations of polarization of various
types of variable stars, studies of interstellar polarization and
even for observations of exoplanets \citep{Berdyugina11}. One of the main applications
has been circular and linear polarimetry of magnetic cataclysmic
variables \citep[e.g.][]{Piirola87,Piirola90,Piirola93,Piirola08}, with orbital periods of a
few hours and rapid variability on time scales from a fraction of a
second to minutes. Simultaneous measurements in the different
passbands are therefore essential.

Another example of a high-precision broadband instrument which was
able to reach an accuracy in stellar optical polarimetry of the
order of few ppm is the {PlanetPol} \citep{Hough}. This instrument
uses PEM as a polarization modulator, three-wedged Wollaston prism as
polarization analyzer and APD as detector. The working frequency of the PEM is
$2f_{0}$ = 40 kHz and it operates as the $\lambda/2$ retarder. To
measure both Stokes parameters of linear polarization, the whole
instrument is rotated by $45^{\circ}$. Moreover, to minimize
systematic errors, the lower section of the instrument with analyzer
and detectors is rotated from $-45^{\circ}$ to $+45^{\circ}$. The
PlanetPol has two independent registration channels, one for the
star (on axis) and one for the sky (off axis). The instrument has
been designed to measure polarization in a wide wavelength range
470--910~nm, centered at the $\lambda = 690$~nm. The effective
wavelength $\lambda_{\rm eff}$ is dependent on the spectral type and
lies in the range of 735--805 nm.

PlanetPol has been mounted in the Cassegrain focus of the 4.2 m WHT
telescope at the ORM (La Palma) and has measured polarization of a
number of bright stars ($m_{V} \leq 5.\!\!^{\rm m}0$) with very
impressive accuracy of few ppm \citep{Bailey10}.

{HIPPI}, a successor to PlanetPol \citep{Bailey15}, employs
the FLC modulator, operated at the frequency of 500 Hz, and PMT
detectors. Unlike PlanetPol, it is optimized for high-precision
polarimetry in the blue wavelengths, in the spectral region $\sim$ 
400--700~nm. Mounted on the 3.9 m Anglo-Australian Telescope, it
was used for high-precision optical polarimetry of the nearby FGK
dwarfs in the brightness range $m_{V} \simeq 0\fm4 - 6\fm0$. The 
achieved accuracy of polarization measurements (depending on brightness) 
is in the range of 1.5--10 ppm \citep{Cotton17}.

{POLISH2} polarimeter uses two PEM modulators operated at
different frequencies to measure three Stokes parameters $q$, $u$
and $v$ in the UBV passbands \citep{Wiktorowicz15}. Reported
accuracy of the instrument for observations of the bright
``naked-eye'' stars with the 3 m  Shane telescope of the Lick
Observatory is at the level of few ppm.

Such an impressive precision of polarimeters with fast modulators
can be only achieved with careful calibration procedure which is
necessary to take into account and minimize all sources of
systematic errors. The accuracy at the level of $10^{-5} - 10^{-6}$
is only possible for bright stars and on telescopes with the
aperture size $\geq 3$ m. Thus, the primary goal of these
instruments is limited to high-precision optical polarimetry of
bright targets. An example of such application is a study of the
interstellar polarization and direction of the local galactic
magnetic field in the vicinity of the Sun \citep{Frisch,Cotton19}. 
The intrinsically non-polarized stars of A, F, and G spectral types 
located at the distance $<$40~pc are bright, but show very low 
($P \leq 0.01\%$ or $ \leq 10^{-4}$) degree of interstellar 
polarization which can only be measured with the instruments 
providing precision at the level of $10^{-5}$ or better.

\subsubsection{High-Speed Broadband Polarimeters for the Optical Wavelengths}
\label{subsubsec:3.1.3} 

Measuring polarization in the optical
wavelengths with time resolution $\le 1$\,s is a challenging task.
An instrument capable to do this must be able to modulate the
polarization of incoming light beam so as to have all Stokes
parameters recorded in a short exposure time, and also to maintain a
high light throughput. This is rather difficult to achieve in
practice. The general approach for fast optical polarimetry is a
continuous rotation of the wave plate or polaroid with the frequency
of few to 10~Hz.

\begin{figure}[t]
\sidecaption
\includegraphics[scale=.50]{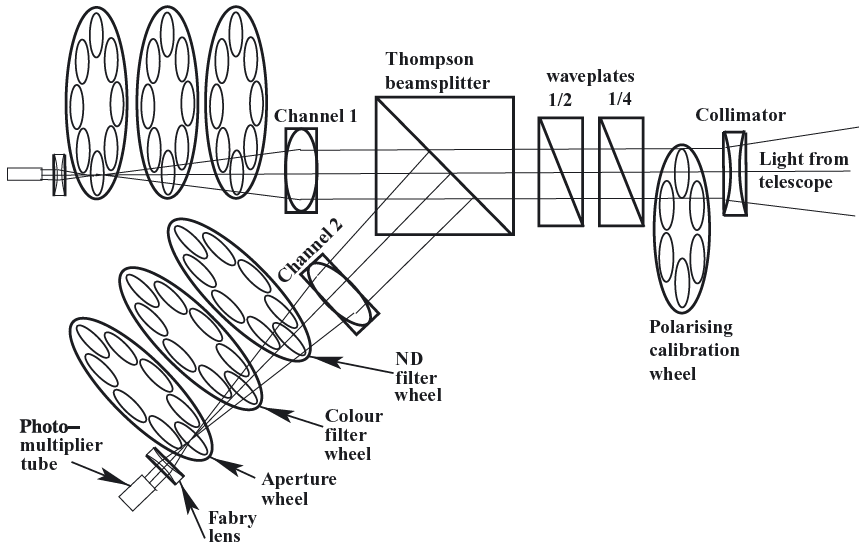}
\caption{The schematic layout of the HIPPO polarimeter. The HWP and
QWP, placed in the collimated beam are contrarotated at frequency of
10 Hz modulating the polarization of the incoming light beam split
by Thomson prism in e- and o-rays and registered separately by the
two PMTs placed in Channels 1 and 2. Linear and circular only modes
of polarimetry are possible and each channel can be used for
independent measurements of three Stokes parameters with different
filters. From \citet{Potter10}.}
\label{fig:8}       
\end{figure}

HIPPO polarimeter, designed in the South African Astronomical
Observatory (SAAO) is an example of an instrument built for
high-speed photo-polarimetry of rapidly varying polarized
astronomical sources, with particular interest in polarimetry of
magnetic cataclysmic variables (mCVs). The layout of the instrument
is shown in Fig.~\ref{fig:8}. HIPPO makes a single polarization
measurement every 0.1\,s \citep{Potter10}. The acquired raw
data, if necessary, can be binned to any integer multiple of 0.1\,s.

Among the results obtained with HIPPO are the discovery of polarized
QPOs in polar IGR J14536$-$5522 \citep{Potter10} and white dwarf
pulsar AR Sco \citep{Potter18}. In both cases, the observations have been done
with the 1.9\,m telescope of the SAAO. The pulse periods, seen also
in brightness variations are 5.2 and $\sim$2\,min,  respectively.

RINGO3\footnote{\url{http://www.telescope.livjm.ac.uk/TelInst/RINGO3/}} mounted on the robotic 2.0\,m Liverpool Telescope at the ORM (La
Palma) employs a rotated polaroid in front of the collimator, after
which the light is split by two dichroic mirrors onto three Andor
iXon 3 EMCCDs. After the year 2013, a Lyot depolarizer has been
installed after the polaroid to get rid of the large instrumental
polarization arising due to the dichroic mirrors. 
Initially the frequency of polaroid rotation was set to 1\,Hz, which has
allowed to measure the linear polarization two times per second.
Currently the rotation frequency is set at $\sim$0.4\,Hz.

The main shortcomings of the RINGO3, apart from rather large
instrumental polarization, are significant overheads for
initializing the instrument for the exposure and the loss of 50\% of
light intensity in the polaroid. Thus, the eight images acquired
during one rotation cycle must be stacked over many cycles in order
to achieve a reasonable S/N. The recommended minimum integration
time is 20\,s \citep{Slowikowska16}.

The GASP polarimeter \citep{Collins13, Kyne16} is a DOAP (division of 
amplitude polarimeter) which uses two EMCCDs with a timing resolution 
down to 500 microseconds. It measures both linear and circular polarization 
simultaneously with no moving parts. 

We note that although the best high-speed polarimeters like HIPPO and 
GASP can sample the polarization with time resolution $\leq$ 0.1\,s, 
the acquired Stokes parameters are usually binned over time 
intervals $\geq 10$\,s to achieve sufficient S/N. Thus, in the case of 
AR Sco, binning over 10\,s has been made \citep{Potter18}. The binning 
interval for the polarization data collected for IGR J14536$-$5522 is 
not mentioned in the original paper, but it can be roughly estimated 
from the plots shown in Figs. 6 and 7 of the paper by \citet{Potter10} 
as $\sim 30 - 60$\,s. Both stars show rather strong (few percent of 
polarization) signal which is strictly periodic and this helps to 
reveal it from the raw data accumulated over the sufficiently long 
time span. Because the accuracy of polarimetry is critically dependent 
on the count rate, the small number of ADUs registered over short time 
interval is generally the problem in detecting weak short-term 
polarization signals. 

It is worth to mention that for time resolution of $\ge$10\,s the
``traditional'' CCD polarimeter with discretely rotated wave plate and
fast readout CCD camera offers comparable, if not better
performance. The single turn of the wave plate is done by stepper
motor in 0.2--0.3\,s. Thus, the time lost in wave plate rotation
can be $\leq 0.5$\,s and $\leq 1.0$\,s for one cycle of circular and
linear polarization measurements, respectively. Stepper motors are
compact, precise and provide reliable long-term operation. Moreover,
modern EMCCD cameras offer frame rates of few tens per second and are 
able to ``stack'' images, thus allowing to shorten the sampling time 
interval considerably.

\subsection{Instruments for Optical Imaging Polarimetry}
\label{subsec:3.2}

\subsubsection{Optical Imaging Polarimetry with Polaroids}
\label{subsec:3.2.1}

The simplest way to measure polarization of extended objects is to
take a series of CCD images with a polaroid filter. Normally, four
images are taken at $0^{\circ}, 45^{\circ}, 90^{\circ}$ and
$145^{\circ}$ orientations of the polaroid. Because the images for
the polarimetry cycle are not taken simultaneously, any variations
in the atmospheric transparency will give spurious polarization
signal. Hence, this method of measurement can be applied only in
strictly photometric nights. The calibration flat field must be
obviously taken separately for each orientation of the polaroid.
This method of imaging polarimetry has been used at the NOT (La
Palma) with the ALFOSC \citep[see, for example,][]{Piirola92,Harjunpaa99}.

\subsubsection{Optical Imaging Polarimetry with a Double-Beam Analyzer and Stripe Mask}
\label{subsec:3.2.2}

The problems with atmospheric transparency variations and light
losses in the polaroid can be solved with the use of a double-beam
analyzer and a \emph{stripe mask}. The FOcal Reducer and low dispersion
Spectrograph (FORS2) mounted on the Cassegrain focus of the 8.2\,m UT1
telescope at the ESO (Chile) is equipped with a mask that can be inserted in the focal plane before the collimator, to avoid overlapping of the two orthogonally polarized beams produced by the Wollaston prism on the CCD. With
this mask, a full scan of the imaging field is achieved by taking
two series of frames displaced by 22\arcsec.\footnote{\url{http://www.eso.org/sci/facilities/paranal/instruments/fors/doc/VLT-MAN-ESO-13100-1543\_v83.pdf}} This allows one to do imaging polarimetry of extended objects and crowded stellar fields.

\begin{figure}[b]
\sidecaption
\includegraphics[scale=.90]{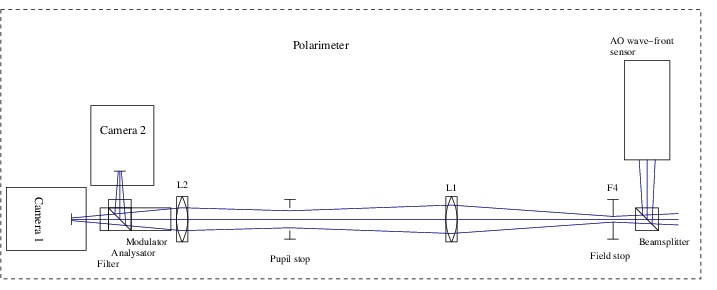}
\caption{The schematic layout of the GPP imaging polarimeter. The
field stop (iris diaphragm) and pupil stop are used to block the
unwanted part of the light. L1 and L2 are collimating and imaging
lenses. After L2 the modulator is installed, followed by PBC unit.
The filters can be inserted either in front of modulator or after
analyzer. From \citet{Gisler16}.}
\label{fig:9}       
\end{figure}

Unfortunately, FORS has a strong field dependent off-axis
instrumental linear polarization which seriously compromises the
performance of the instrument for multi-target or extended object
polarimetry.\footnote{\url{http://www.eso/org/sci/facilities/paranal/instruments/fors/inst/pola.html}} 
Hence, this instrument is mostly used for
broadband and spectropolarimetry of single targets in the center of
the field of view.

\subsubsection{Optical Imaging Polarimetry with Fast-Frequency Modulators and Fast-Readout CCDs}
\label{subsec:3.2.3}

The idea of using fast-readout EMCCD cameras in combination with the
FLC modulators for optical imaging polarimetry is attractive. An
example of this approach is the GREGOR Planet Polarimeter \citep[GPP,][]{Gisler16}. This design eliminates the necessity to use the stripe mask and allows recording of the two orthogonally polarized images
simultaneously and synchronously with the fast modulation and two
EMCCD cameras.

The layout of GPP is shown in Fig.~\ref{fig:9}. GREGOR is a 1.5 m
solar telescope located at the Observatorio del Teide, Tenerife.
This telescope, which is a double Gregory-type design, can be also
employed for night-time observations. The polarimeter is located on
the stationary optical bench after the 50/50 cube beam splitter
which reflects part of the light to the wave front sensor. The
transmitted light is used by the polarimeter. The camera lens (see
Fig.~\ref{fig:10}) images the focal plane to the CCD cameras with a
resolution of $0\farcs086$/pixel. The polarization unit consists of
modulator and analyzer. The modulator unit has two FLCs and two
wave plates which allows to measure all Stokes parameters. A
polarizing beam-splitter cube (PBC) is used as the analyzer. Two
interchangeable modulators optimized for the the blue-visible and
visible - read spectral regions can be used. The modulation
frequency is determined by the ANDOR iXon Ultra 897 camera frame
rates. For the full frame images it is 14 Hz. For the sub-frame
imaging mode, e.g. used for polarimetry of Uranus and Neptune, with
the frame size of 128$\times$128 pixels, the modulation frequency is 150 Hz \citep[see][]{Gisler16}.

The polarimeter can measure all Stokes parameters either with a single
or two CCD cameras at the same time. Due to multiple reflections in
telescope mirrors and alt-azimuth mount, the telescope instrumental
polarization is high and variable. To account for it, GREGOR has a
special polarimetric calibrating unit installed in the telescope
before the first diagonal mirror. The calibration procedure consists
of taking calibration measurements at the same position as the
object observed at the sky. Currently only relative calibration of
polarization within the image is possible, but the artificial
unpolarized flat field light source will be implemented in the
future. This should allow an absolute calibration of polarization
image.

\subsection{Instruments for Optical Spectropolarimetry}
\label{subsec:3.3}

In many cases, spectropolarimetry in the optical wavelengths can be
done by adding the polarization modulator and analyzer to the
optical path of the spectrograph. This method is well suitable for
low and intermediate resolution spectropolarimetry and has been
implemented in many existing instruments (see Section \ref{subsubsec:3.1.1}). The processes of data acquisition and data reduction do not differ much
from those applied for traditional spectroscopy where only
$I(\lambda)$ is registered (Section  \ref{subsubsec:2.4.1}.)

\subsubsection{Instruments for Low and Mid-Resolution Optical Spectropolarimetry}
\label{subsec:3.3.1}

One example of a multi-functional instrument with spectropolarimetry
observation mode implemented is FORS2. The polarization unit consists of a
discretely rotated wave plate (HWP or QWP), placed after the collimator into the parallel beam with the dedicated swing arm, and Wollaston prism, mounted in the uppermost filter wheel. With the inserted polarization optics, spectropolarimetry can be done with most of the available grisms, with the spectral resolution $\lambda/\Delta\lambda$ from 450 to 2500. FORS2 employs superachromatic wave plate mosaics ($3\times3$) and can be used for single and multi-object spectropolarimetry in the spectral range 330--1100 nm.

Spectropolarimetry can be done with the two CCD mosaics, either with
the two $2k\times4k$ MIT CCDs optimized for the red part of spectrum
or with two $2k\times4k$ E2V CCDs particularly sensitive in the blue
range. Because FORS2 is mounted in the Cassegrain focus, it benefits
from the very low telescope polarization (in the center of the field
of view) and thus is well suitable for the absolute linear and
circular spectropolarimetry of faint objects, being one of the best
instruments in this class. FORS1 (a retired twin of FORS2) has been
extensively used for the study of stellar magnetic fields with
spectropolarimetry  \citep[see, for example,][]{Bagnulo02,Bagnulo06,Vornanen10}.

Spectropolarimetry with ALFOSC on the 2.52 m Nordic Optical Telescope (NOT) at the ORM (La Palma) is done with the HWP or QWP plates mounted in the converging beam (before the focal plane) and calcite beam-splitter mounted in the aperture wheel. The spectral resolution in the wavelength range 320--1100 nm can be selected from 210 (with the low-res, wide spectral range grism) up to 10000 (high-res grism, centered at the H$\alpha$).

\begin{figure}[b]
\includegraphics[scale=.14]{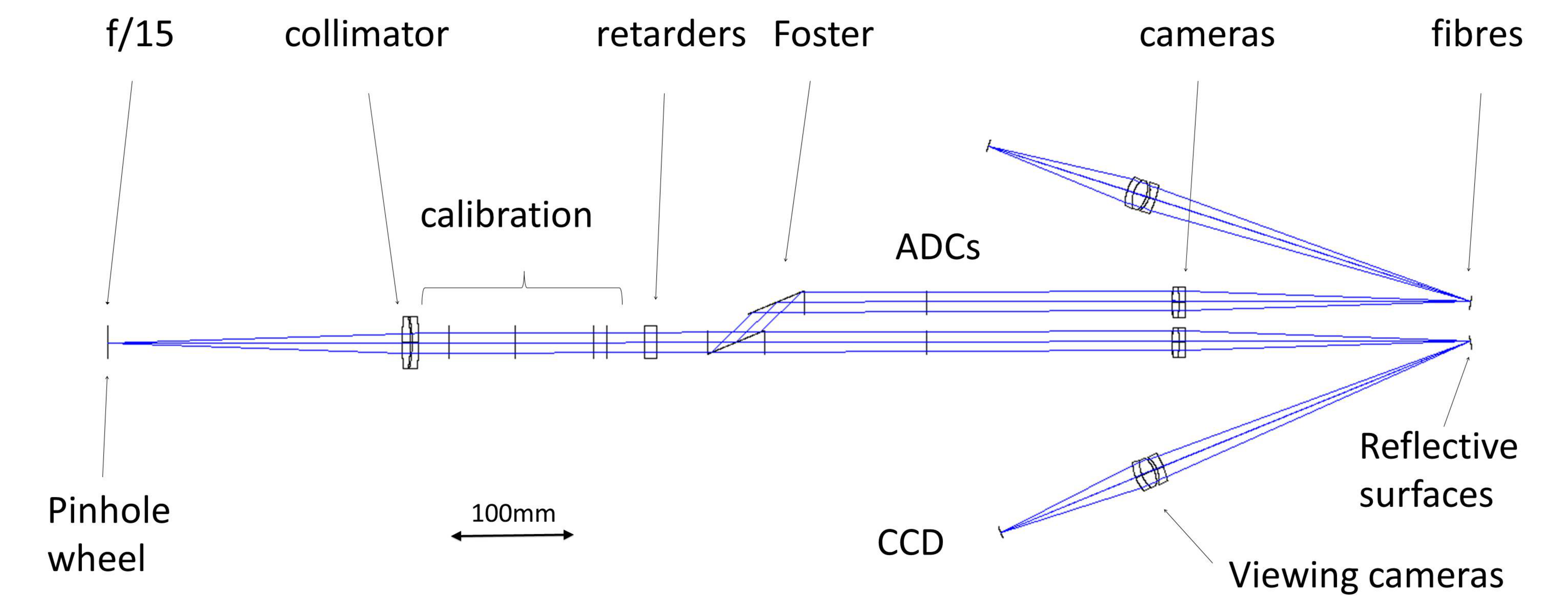}
\caption{Optical design of the PEPSI spectropolarimeter. The light
enters from the left and exit into the fibers to the spectrograph
section on right. The fibers are positioned on the central pinholes
in the reflective surfaces. The individual optical components are
marked. ADCs are the atmospheric dispersion correctors. The
calibration components and QWP unit are retractable. From
\citet{Strassmeier15}.}
\label{fig:10}       
\end{figure}

\subsubsection{Instruments for High-Resolution Optical Spectropolarimetry}
\label{subsec:3.3.2}

The most important application of high-resolution spectropolarimetry is 
for study of magnetic fields in solar-type stars. Spectral resolution higher 
than $\sim50000$ normally requires an \'{e}chelle spectrograph.

Good example of a polarimeter implemented to a high-resolution
\'{e}chelle spectrograph is PEPSI (Potsdam Echelle Polarimetric and
Spectroscopic Instrument) built for the 8.2 m Large Binocular
Telescope (LBT). The design allows to measure all four Stokes
parameters ($I$, $Q$, $U$ and $V$) with two identical independent
polarimeters at the spectral resolution of 130000 \citep{Strassmeier15}. These
polarimeters are installed at the straight-through f/15 Gregorian
foci of the LBT. The entire PEPSI light feed is split into the
\emph{integral light} part with the permanent focus and
\emph{polarimetric-light} part which is available after dismounting
other instruments \citep[see][for further details]{Strassmeier15}.

The polarimeter units are of double-beam type with a Foster prism
beam-splitter used as the polarization analyzer. For measuring
linear polarization the analyzer unit is rotated. For measuring
circular polarization, a superachromatic QWP is inserted into the
light beam. The principal layout of the instrument is shown in
Fig.~\ref{fig:10}. Detailed description of the instrument can be also
found in \citet{Ilyin11}. Spectropolarimetry with high-resolution spectrographs
which always employ a significant number of reflecting optical
elements, \'{e}chelle grating, multiple mirrors, fibres etc.,
requires careful calibration of the instrumental polarization 
\citep[e.g.][]{Ilyin12}. The calibration process involves modeling the transformation of the Stokes parameters in the telescope and polarimeter optics
with the Mueller matrices. The precise calibration is normally done
with the use of polarized calibration light sources inserted into
the optical path.

\section{Calibration Techniques for Optical Polarimetry}
\label{sec:4}

The advantage of polarimetry over photometry is that the measurement
is based on the \emph{relative} intensity of two perpendicularly
polarized components of light, rather than the total intensity
measured in photometry. That helps in eliminating effects from
variable atmospheric transmission and the throughput of the
instrument. However, on the counterside is the very minute amount of
polarization seen in astronomical objects, rarely exceeding a few
percent and in the most demanding applications detection sensitivity
at the ppm level ($10^{-6}$) is required.

\subsection{Calibration of Linear Polarization}
\label{subsec:4.1}

In case of linear polarization measurements, the calibration has to
be carried out for:
\begin{enumerate}
\item{Polarization scale,}
\item{Zero-point of polarization angle,}
\item{Instrumental polarization produced by the polarimeter,}
\item{Polarization produced by the telescope optics. }
\end{enumerate}
A straightforward way of polarization scale calibration (1) is to
insert a high quality polarizer, e.g. a Glan prism with a nearly
100\% efficiency, in front of the polarimeter. The scale calibration
is particularly important for the high frequency modulators, where
the efficiency can be as low as $\sim$ 70\% and strongly dependent 
on the wavelength, as the modulators are not achromatic. Double image
polarizers, like calcite plate and Wollaston prism, have practically
100\% efficiency. If the whole polarimeter is rotated at
$45^{\circ}$ steps, virtually no polarization scale calibration is
then needed. Superachromatic retarder plates with excellent
performance in the 310--1100 nm range are available, but scale
correction factors up to 1.02--1.03 may be encountered also with
these devices. In practice, if no scale calibration optical
component is available, one has to trust on published high
polarization standards to correct for the polarimeter efficiency
factor.

For the angle calibration (2) one may use a high quality polarizer
oriented accurately in a known, e.g. equatorial, frame of
references. The double image polarizers, calcite and Wollaston
prism, are particularly advantageous in this regard, because they
provide images with known polarization orientation, readily fixed in
the celestial coordinate system by the line connecting the two
images. For that reason a rotatable ($45^{\circ}$ steps) double
image polarimeter is an ideal instrument for establishing high
polarization standards, both for the scale and the angle
calibration. Superachromatic retarder plates are less favourable for
the highest precision angle calibration, because their optical axis
typically varies in a complex way by several degrees over the
wavelength regions applied. Therefore, it is also crucial not
to use too broad passbands ($<$ 100 nm) if the angle calibration is
critical. Differences in spectral energy distribution could easily
introduce noticeable shifts in the effective wavelength and thereby
errors in the corrections applied.

The best way of eliminating the instrumental polarization (3)
produced by the polarimeter ($q_{\rm in}$, $u_{\rm in}$) is to rotate the
whole instrument through $360^{\circ}$ by $45^{\circ}$ steps for one
complete measurement. Typically, rotation effects (optics
misalignments etc.) create a sinusoidal spurious modulation, but
since the stellar polarization gives a \emph{double} sinusoid, the
instrumental effects are largely cancelled in the reductions.
Rotating the whole polarimeter, however, is often undesirable and a
rotated retarder is used instead. Similarly, the half-wave plate
must be rotated full $360^{\circ}$ cycles ($22\fdg5$ steps) to
get the best accuracy and avoid effects from e.g. dust particles on
the retarder, non-parallelism of rotating components, etc.

Eliminating the spurious polarization produced by the telescope
optics (4) is perhaps the most difficult part to establish without
leaving residual systematic errors, if the highest precision (a few
ppm level) is required. Such a source like an ``unpolarized star''
probably does not exist. Stellar chromospheric activity may lead to
detectable, and variable, intrinsic polarization even for ``normal''
A-G main sequence stars. For a reliable determination it is
necessary to observe a sample of (5 -- 20) nearby stars ($d < 25$ pc,
if possible) and obtain the normalized Stokes parameters of the
telescope polarization, ($q_{\rm tel}$, $u_{\rm tel}$) as the average of the
$(q, u)$ of the observed sample of stars. In this way also small
effects of interstellar polarization present in each of the observed
stars will tend to cancel out.

Alt-az telescopes provide an additional way of determining
($q_{\rm tel}$, $u_{\rm tel}$). Because the field angle of the telescope
optics continuously changes on the sky, this gives polarization
modulation: a double cosine curve over one full rotation. Fitting
curve to this modulation yields the amplitude and angle of the
telescope polarization in the telescope optics coordinate system.
These can then be rotated to the equatorial frame of references for
the moment of the observation and subtracted from the stellar
observation.

There are numerous ways, in addition to curve fitting, to determine
the ($q_{\rm tel}$, $u_{\rm tel}$) for an alt-az telescope. During
observations the polarimeter is rotated with respect to the
telescope to keep it in the equatorial system. The observed $(q, u)$
of zero-polarization standard stars can be rotationally transformed
to the telescope optics system. After the transformation, the
($q_{\rm tel}$, $u_{\rm tel}$) values are obtained by averaging the $(q, u)$
values of the standard stars in the telescope frame of references.
Also iterative methods can be applied to get both ($q_{\rm in}$,
$u_{\rm in}$) and ($q_{\rm tel}$, $u_{\rm tel}$) from the same sample of
calibration measurements.

Equatorial mounted telescope is convenient in the sense that the
telescope polarization does not rotate on the sky. It only gives a
constant shift in $(q, u)$. This is particularly important for
searching small short-term periodic variations in binary and
multiple stars, exoplanet systems, etc. Any residual telescope
polarization does not change the shape of the observed phase-locked
patterns of the normalized Stokes parameters $(q, u)$. The curves
are only shifted by the amount of uncorrected ($q_{\rm  tel}$,
$u_{\rm tel}$), similarly to the effect of interstellar polarization.

\subsection{Calibration of Circular Polarization}
\label{subsec:4.2}

The telescope \emph{circular} polarization ($v_{\rm tel}$) is of an
order of magnitude smaller than linear and, therefore, can be
ignored in many applications. However, the effect of
\emph{cross-talk}, i. e. the transformation of high linear polarization into circular which may occur in the telescope or
polarimeter optics, must be carefully investigated and taken into
account. The sign of the circular polarization can be calibrated
with the observations of the magnetic white dwarfs with the strong
and non-variable circular polarization, e.g. Grw+70 8247 which has
$P_{\rm C}\simeq - 4.0 \%$ in the B-band.

\subsection{Calibration of the CCD Polarimetry: Flatfielding}
\label{subsec:4.3}

Standard CCD reductions, bias, dark, and flat field calibrations are
obviously applied in the first part of the reduction pipeline
scripts. Our experience with double image polarimeters having a
rotatable superachromatic retarder (DiPol-1 and DiPol-2), and
spectropolarimeters (ALFOSC at the NOT, EFOSC and FORS at the ESO),
is that it is best not to apply different flat fields for each of
the different position angle of the retarder, but use the same flat
field for all (16) retarder positions. Individual flats fail to
improve the precision, on the contrary. Common flat field is enough
to suppress pixel-to-pixel variations in the CCD sensitivity.
Measuring sequences consisting of cycles made over full rotation of
the retarder make the polarimeter rather insensitive to flat field
imperfections.

ESO staff recommends to take sky flats for calibration of polarimetry
images acquired with the FORS with the \emph{polarization optics
removed}. The reason for this is that the sky flat itself can be
polarized. To avoid possible errors, it is advisable always to take
sky flats in the direction of sky opposite to the rising or setting
Sun.

For the highest S/N measurements it is advantageous to strongly
defocus images to spread light over a very large number of pixels 
\citep{Piirola14}. In this way we can expose up to $10^{8}$ electrons 
in one stellar image without saturating the CCD pixels. Even if there 
are minor shifts in the position of the two perpendicularly polarized 
images of the target, vast majority of the pixels remain the same over 
the full measurement cycle (16 exposures). Minor imperfections in the
flat field are eliminated in the reductions, since the ratio of the
o- and e-beam transmission and efficiency, if constant, is
automatically cancelled in the reduction algorithm. This provides
inherently very stable instrument and detection sensitivity better
than $10^{-5}$ (< 10 ppm) in $\sim$1 hour for sufficiently bright
stars. In fact, e.g. DiPol-2 polarimeter is photon-noise limited down
to these very low polarization signal levels.   

\section{High-precision CCD polarimetry of binary stellar systems}

One important application of optical polarimetry is to study  
interacting binary stars. The interaction, which involves transfer
of matter from one component to another, very often gives rise to
variable linear polarization due to the light scattering on the 
circumstellar material such as gaseous streams, discs, jets and 
non-spherical envelopes. In binary systems with accreting 
black hole component, linear polarization may also arise due to 
synchrotron radiation emitted in the jet. Thus, polarimetry of 
these objects can provide useful information on their physical 
properties.
Below we provide examples of high-precision measurements of  polarization in black hole X-ray binaries and in early type binaries, made with the DiPol-2 instrument. 

\begin{figure}[b]
\begin{minipage}{0.48\columnwidth}
\includegraphics[width=\columnwidth,height=95mm]{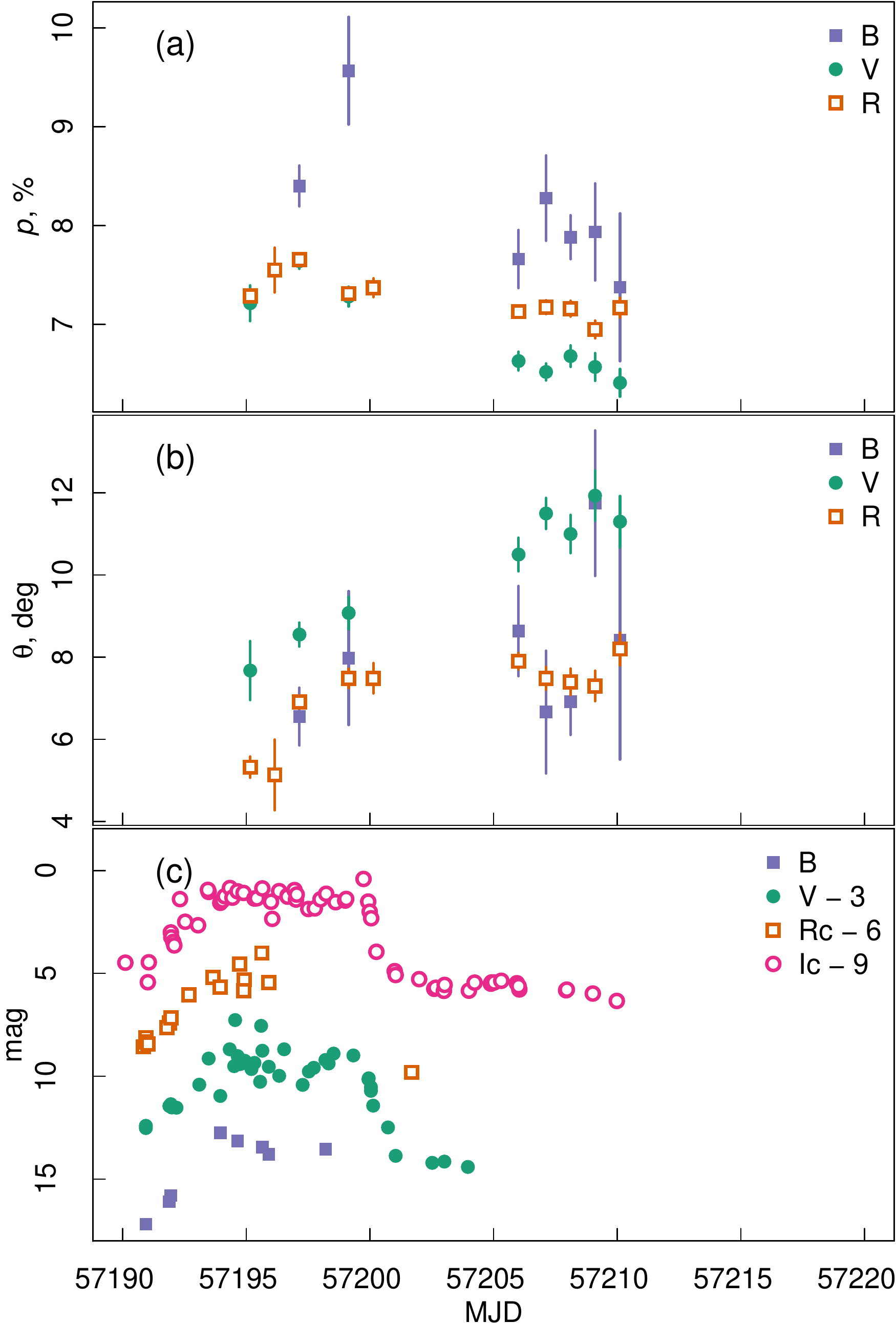}
\end{minipage}
\begin{minipage}{0.48\columnwidth}
\includegraphics[width=\columnwidth,height=45mm]{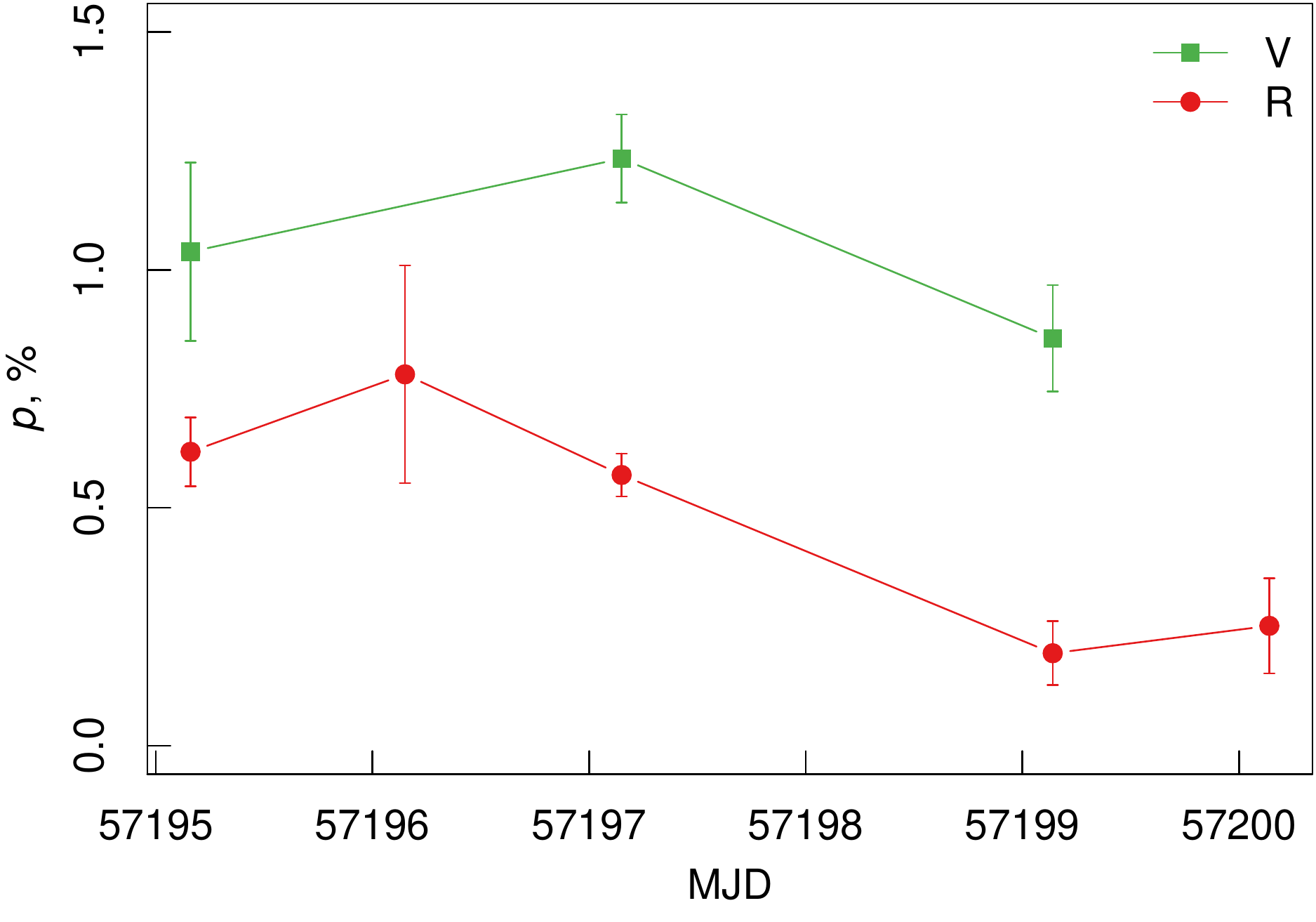}
\\[3mm]
\includegraphics[width=\columnwidth,height=45mm]{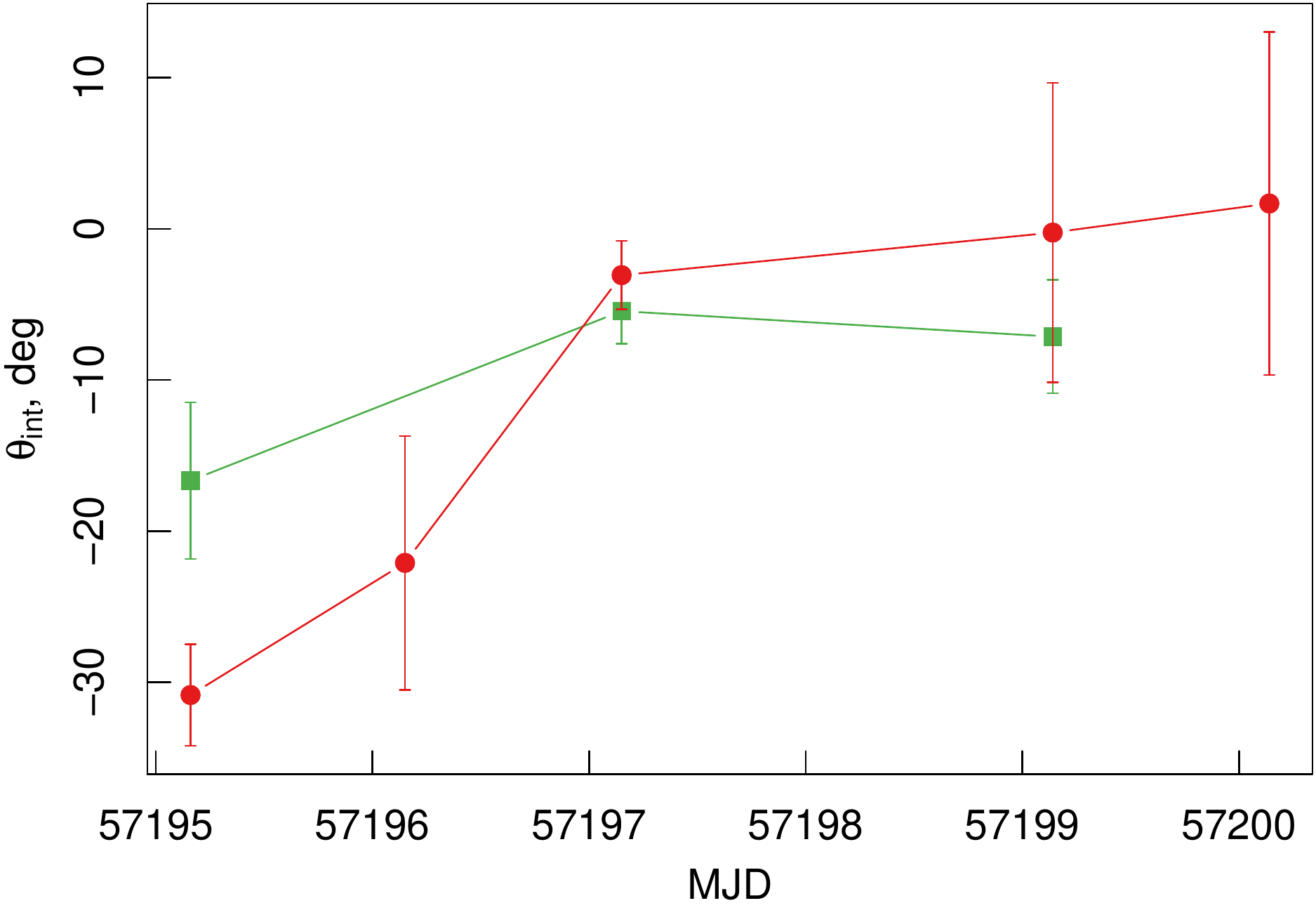}
\end{minipage}
\caption{The 2015 June--July outburst of the black hole X-ray binary V404 Cyg. \emph{Left}: panels (a) and (b) show the observed linear PD ($p$) and PA ($\theta$) in three bands; and panel (c) gives the optical light curve. 
\emph{Right}: variations of the intrinsic linear PD ($p$) and PA ($\theta$) in the V and R bands over the outburst. From \citet{Kosenkov}.}
\label{fig:11}       
\end{figure}
 
\subsection{Polarimetry of black hole X-ray binaries}
\label{subsec:blackholes}
 
The low-mass black hole X-ray binary V404 Cyg went into outburst in 2015 after being 26 years in quiescence. 
In the peak of the outburst, the object showed very erratic behaviour with flares reaching a few tens of Crab in the hard X-ray domain, which for the known distance of 2.4 kpc corresponds to the Eddington luminosity for a  10\,${M}_{\odot}$ black hole \citep{Rodriguez15,Motta17}.   
The object also showed strong variability in the optical \citep[see][and also lower left panel of Fig.~\ref{fig:11}]{Kimura2016}.  
In order to understand the nature of the optical emission we performed simultaneous three-colour (BVR) polarimetric observations during and after the outburst \citep[see][for details]{Kosenkov}. 
We detected small but statistically significant change of the linear PD  by $\sim$1\% between the outburst and the quiescence (see left panels of Fig.~\ref{fig:11}). 
We also found that the polarization of V404 Cyg in the quiescent state agrees well with that of the visually close (1\farcs4) companion, as well as those of the surrounding field stars, indicating that it is predominantly of interstellar origin. 
From the observed variable polarization during the outburst we showed that the intrinsic polarization component peaks in the V band with  PD$_{\rm V} = 1.1\pm 0.1$\%, and in the R band it is factor of two smaller PD$_{\rm R} = 0.46\pm 0.04$\%, while the polarization PA  ($\theta$) = $-7^\circ\pm 2^\circ$ is similar in all three passbands. 
The observed wavelength dependence of the intrinsic polarization does not support non-thermal synchrotron emission from a jet as a plausible mechanism, but is in better agreement with the combined effect of electron (Thomson) scattering and absorption in a flattened plasma envelope or outflow surrounding the illuminating source. 
Interestingly, we also found that the PA of the intrinsic polarization is nearly parallel to the jet direction (i.e. perpendicular to the accretion disc plane) as determined by the VLBI observations. 

\begin{figure}[h!]
\begin{minipage}{0.50\columnwidth}
\includegraphics[width=\columnwidth,height=130mm]{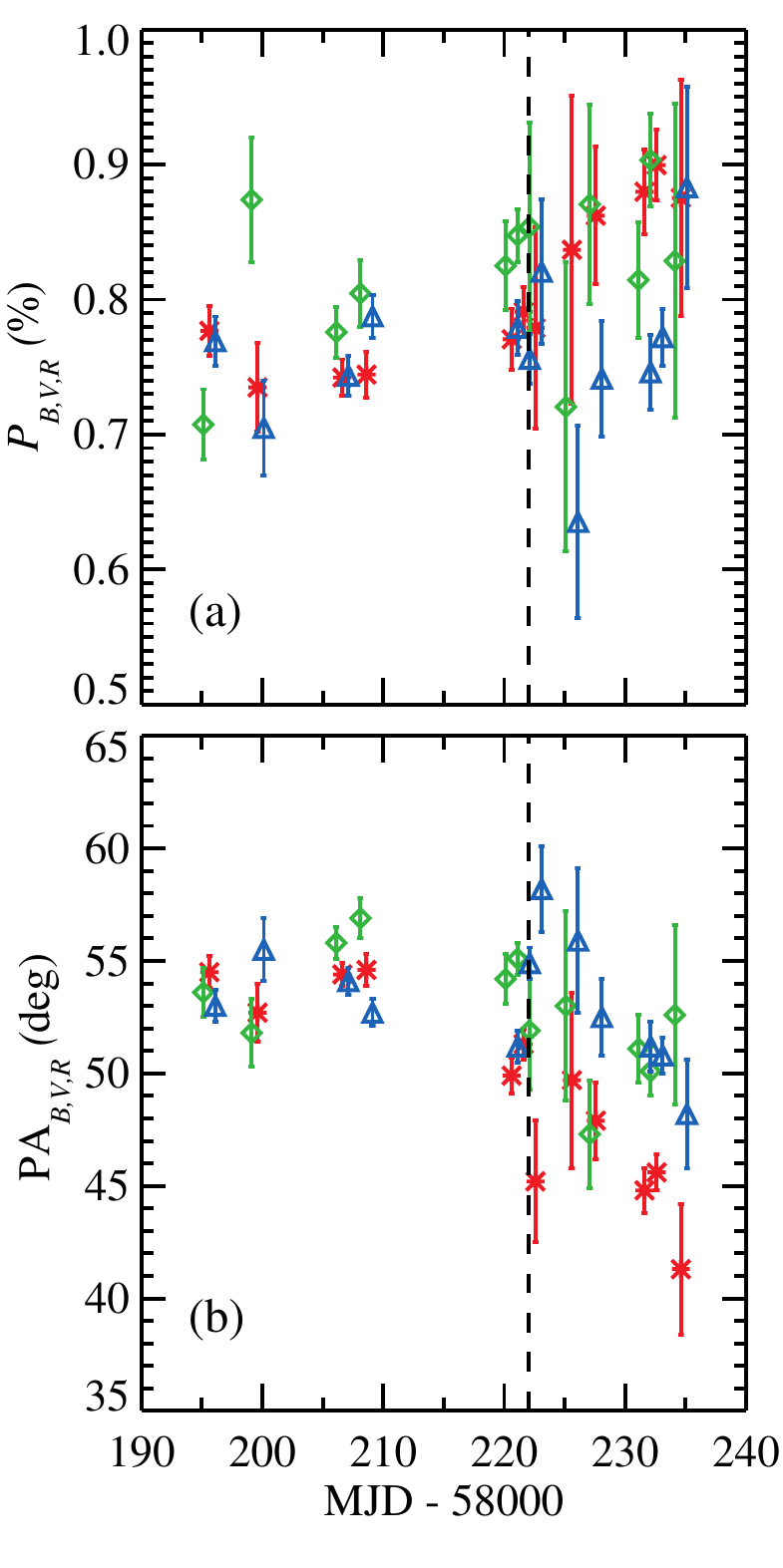}
\end{minipage}
\begin{minipage}{0.50\columnwidth}
\hspace{-0.2cm}
\includegraphics[width=0.9\columnwidth,height=45mm]{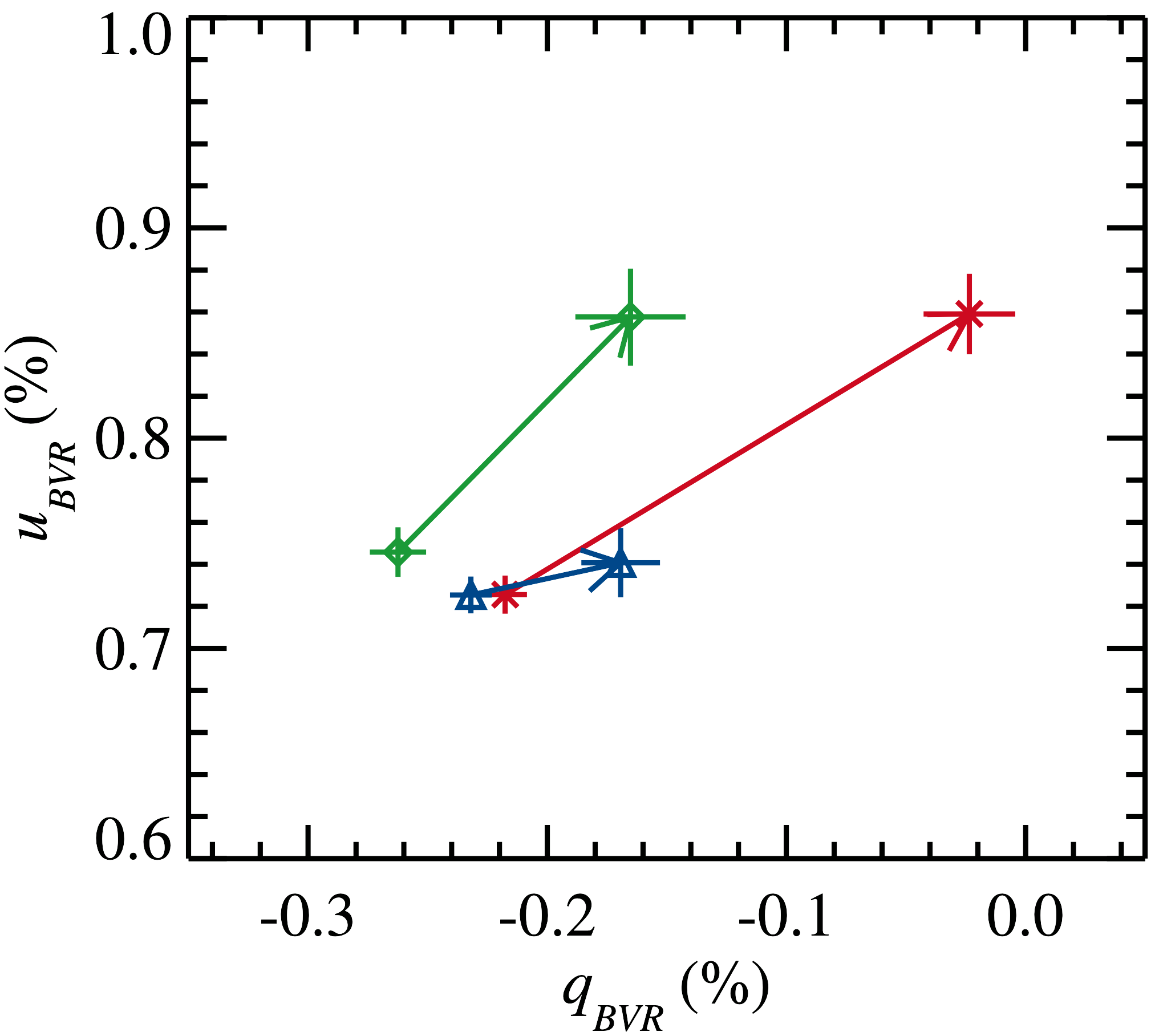}
\\ 
\hspace{0.cm}
\includegraphics[width=0.9\columnwidth,height=75mm]{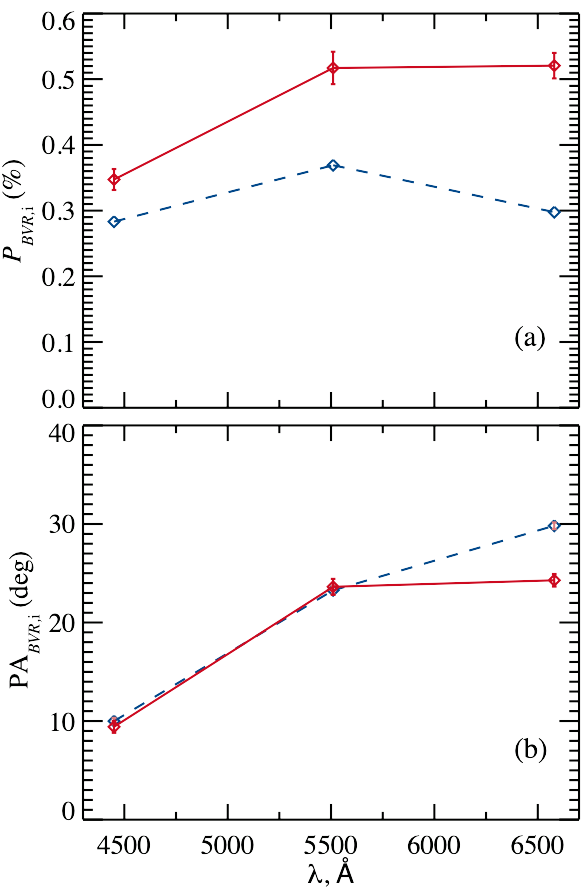}
\end{minipage}
\caption{Polarization properties of black hole X-ray binary MAXI J1820+070. 
Left panels: Evolution of (a) PD ($P$) and (b) PA in different filters: B (blue triangles), V (green diamonds) and R (red crosses).  
The vertical dashed line marks MJD~58222 (2018 April 14), after which we detected a significant increase of polarization. 
Right upper panel: Observed polarization in three filters (B - blue triangles, V - green diamonds and R - red crosses) in the Stokes $(u,q)$-parameter plane before (lower left symbols) and after MJD 58222 (upper right symbols). 
Arrows show the direction of polarization change. 
Right two lower panels: Wavelength dependence of the \emph{intrinsic} (a) PD ($P$) and (b) PA. 
Weighted averages are shown with the blue dashed and red solid lines for observations before and after MJD~58222, respectively. 
Adapted from \citet{Veledina}.}
\label{fig:12}       
\end{figure}

Another black hole X-ray transient MAXI~J1820+070 went into a spectacular outburst in March 2018.  
In the X-rays the source flux exceeded 3~Crabs and in the optical reached magnitude $m_{V}=$12--13.  
This unusually bright event allowed detailed investigation of multiwavelength spectral and timing properties and the source was monitored from radio to the $\gamma$-rays.
Fast variability and powerful flares in the optical and infrared as well as optical and X-ray quasi-periodic oscillations were detected in the source.
To disentangle sources of optical emission we performed BVR polarization measurements during March--April 2018 \citep[see][for details]{Veledina}. 
We detected small, $\sim$0.7--0.9\%, but statistically significant linear polarization at PA of about 50$^\circ$ in all filters.
We also detected  a significant change in  the PD after 2018 April 14 (see right panels in Fig.~\ref{fig:12}). 
The change is of the order of 0.1\% and is most pronounced in the R band.
By analyzing a set of nearby field stars, we were able to determine the contribution of the interstellar polarization. 
This allowed us to obtain the intrinsic polarization of MAXI~J1820+070 at the level of 0.3--0.5\%, depending on the filter (see middle right panel in Fig.~\ref{fig:12}). 
The change of the source Stokes vector occurred simultaneously with the drop of the observed V magnitude and a slow softening of the X-ray spectrum, which we interpreted as a signature of an evolution towards the soft state.  
The increased polarization may be a result of the decreasing contribution of the non-polarized component which we can associate with the hot flow or a jet. 
Its low polarization likely results from the tangled geometry of the magnetic field or from the Faraday rotation in the dense, ionized and magnetized medium close to the black hole. 
We argued that the polarized optical emission is produced by the irradiated disc or scattering of its radiation in the optically thin outflow.

\subsection{Polarimetry of early type binaries}
\label{subsec:earlytype}
 
\begin{figure}[h!]
\includegraphics[scale=.55]{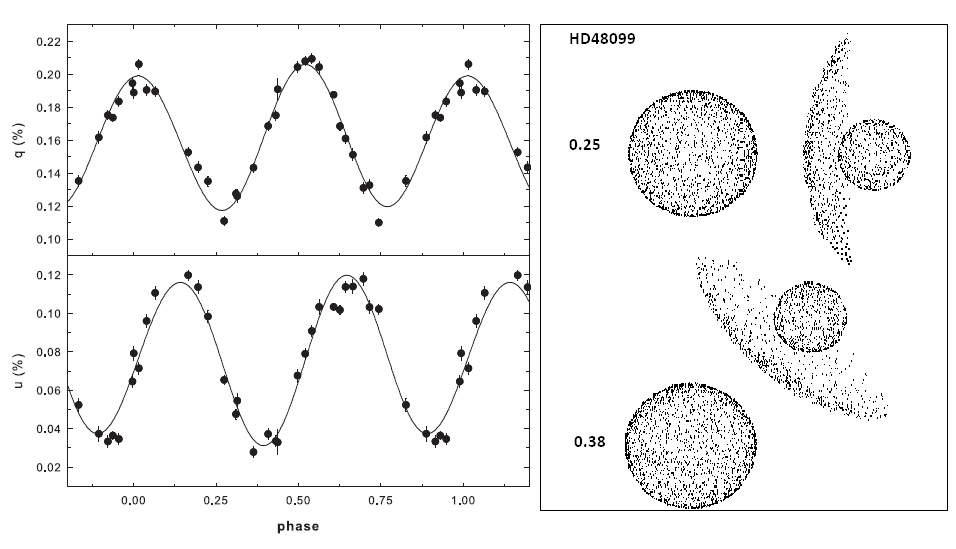}
\caption{Polarization variability in HD48099. \emph{Left}: Observed 
 Stokes parameters $q$ and $u$ in the B-band, plotted over the phase 
of the orbital period. The vertical bars show $2\sigma$ errors. \emph{Right}: The scattering model showing O-type components 
and the light scattering cloud produced by colliding stellar winds at the different phases of orbital period. Adapted from \citet{Berdyugin16}.}
\label{fig:13}    
\end{figure}

\begin{figure}[t]
\includegraphics[scale=.45]{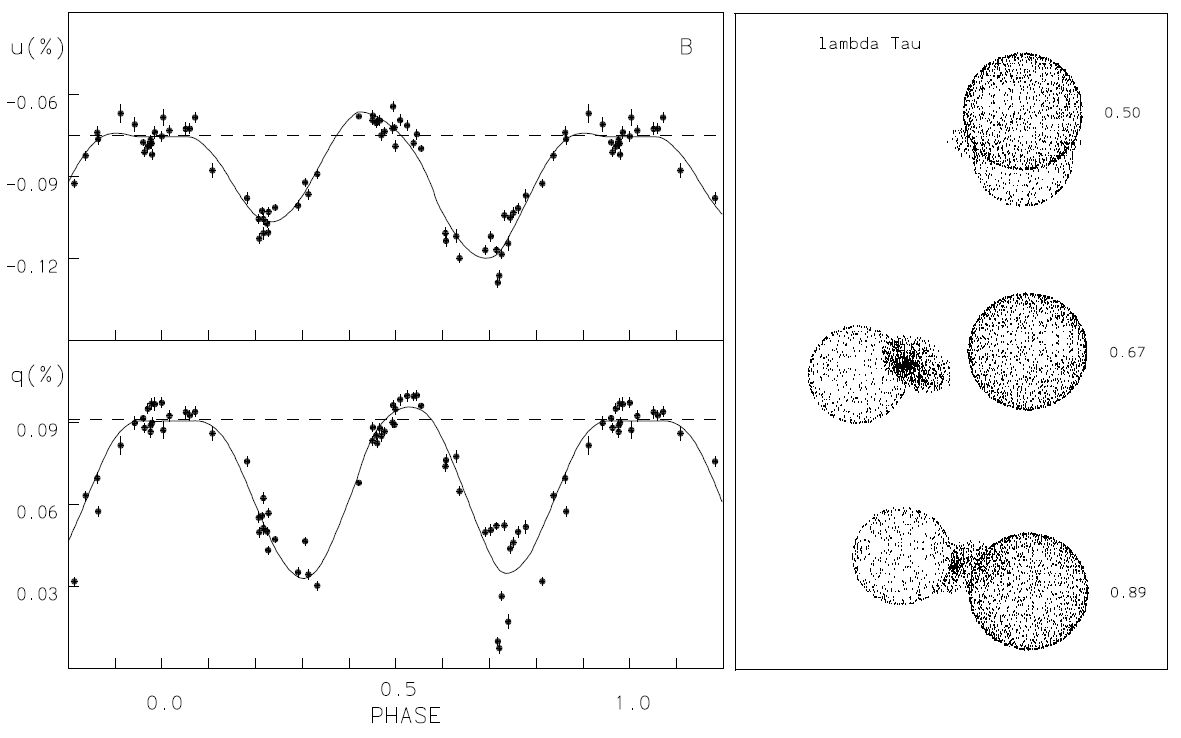}
\caption{Polarization variability in $\lambda$ Tauri. \emph{Left}: Observed Stokes parameters $q$ and $u$ in the B-band plotted over the phase 
of the orbital period. The vertical bars show $2\sigma$ errors. \emph{Right}: The scattering model showing primary and secondary components with the light scattering cloud between them at the different 
phases of orbital period. Adapted from \citet{Berdyugin18}.}
\label{fig:14}       
\end{figure}

In massive early-type binaries and various interacting systems (e.g. Algol-type), orbitally phase-locked linear polarization may arise due to electron (Thomson) scattering. A detailed analysis of the polarization variability yields independent estimates of the orbital orientation $\Omega$ and inclination $i$ \citep[e.g.][]{Brown78}. The amplitude of the variability depends on the amount of the light-scattering material and usually is $\leq 10^{-3}$. In order to detect this variability, a high precision, at the level of $10^{-4}$ or better, is necessary. Examples of applications of CCD polarimetry on such binary systems are given in Figs. \ref{fig:13} and \ref{fig:14}. 
 
 HD48099 (Fig. \ref{fig:13}) is a massive, detached early-type binary consisting of O5.5\,V and O9\,V stars. The orbital period is $\sim$3.08 day. 
 There is evidence of colliding stellar winds the system \citep[e.g.][]{Mahy10}.
 This binary is non-eclipsing, and an independent estimate of the orbital inclination, provided by polarimetry, is important for constraining the component masses. 
 
In Fig. \ref{fig:14} we show our results for $\lambda$ Tau, a ``classic'' Algol-type triple system with the inner binary consisting of a more evolved and less massive mid-A type secondary filling its Roche lobe and a more massive B3\,V primary star. 
The orbital period of the inner pair is $\sim$3.95 days. The unseen tertiary star (most probably a K dwarf) orbits around the inner binary with the period of $\sim$33 days \citep[e.g.][]{Fekel82}. 
 
In both cases, small-amplitude variations of polarization ($\sim$0.1\% for HD48099 and $\sim$0.05\% for $\lambda$ Tau) have been revealed with DiPol-2 polarimetry. Modelling the polarization data yielded the determination of the orbital parameters and the location of light scattering material \citep{Berdyugin16,Berdyugin18}. 
 
\section{Summary}
\label{sec:summary}
 
In this chapter we reviewed the methods, instruments and calibration techniques used in modern astronomical optical polarimetry. 
We described the properties of various polarization devices and detectors used for optical broadband, imaging and spectropolarimetry, and discussed
their advantages and disadvantages. 
We stressed the necessity to properly calibrate the raw polarization data, and discussed the methods of the determination of instrumental polarization and its subtraction.
Finally, we presented a few examples of high-precision measurements of optical polarization made with our DiPol-2 polarimeter, which allowed us to obtain interesting constraints on the origin of optical emission in black hole X-ray binaries and on the orbital parameters of massive stellar binaries. 

\section*{Acknowledgements}

We acknowledge support from the ERC Advanced Grant HotMol ERC-2011-AdG-291659 (AVB).
The DiPol-2 was built in cooperation between the University of Turku, Finland, and the Kiepenheuer Institut f\"ur Sonnenphysik, Germany, with support from the Leibniz Association grant SAW-2011-KIS-7.

 \bibliographystyle{spbasic}
  \bibliography{ref_polar}

\end{document}